\newcommand{\NPA}[3]{Nucl.\ Phys.\ A\ {\bf #1},\ #2 (#3)}
\newcommand{\NPB}[3]{Nucl.\ Phys.\ B\ {\bf #1},\ #2 (#3)}

\newcommand{\PLB}[3]{Phys.\ Lett.\ B\ {\bf #1},\ #2 (#3)}
\newcommand{\PR}[3]{Phys.\ Rep.\ {\bf #1},\ #2 (#3)}
\newcommand{\PRL}[3]{Phys.\ Rev.\ Lett.\ {\bf #1},\ #2 (#3)}

\newcommand{\PRC}[3]{Phys.\ Rev.\ C\ {\bf #1},\ #2 (#3)}
\newcommand{\PRD}[3]{Phys.\ Rev.\ D\ {\bf #1},\ #2 (#3)}

\newcommand{\EPJC}[3]{Eur.\ Phys.\ J.\ C\ {\bf #1},\ #2 (#3)}
\newcommand{\EPJA}[3]{Eur.\ Phys.\ J.\ A\ {\bf #1},\ #2 (#3)}

\newcommand\g{\gamma}

\newcommand\e{\epsilon}

\renewcommand\i{\iota}

\newcommand{\diracslash}[1]{#1\llap{/\kern2pt}}

\newcommand{\be}{\begin{equation}}
\newcommand{\ee}{\end{equation}}
\newcommand{\bea}{\begin{eqnarray}}
\newcommand{\eea}{\end{eqnarray}}
\newcommand{\ba}[1]{\begin{array}{#1}}
\newcommand{\ea}{\end{array}}

\documentclass[prd,aps,floats,nofootinbib,tightenlines,showpacs]{revtex4-1} %[twocolumn,10pt]{revtex4-1}
\usepackage{epsfig,graphicx,pstricks}
\usepackage{psfrag}
\usepackage{color}
\usepackage{amsmath}
\usepackage{amsfonts}
\usepackage{amssymb}
\usepackage{textcomp}
\usepackage{multirow}
\usepackage{float}

\begin{document}

\title { Transport coefficients in the Polyakov quark meson coupling model:
 A relaxation time approximation}
\author{Aman Abhishek }
\email{aman@prl.res.in}
\affiliation{Theory Division, Physical Research Laboratory,
Navrangpura, Ahmedabad 380 009, India}
%\author{Pracheta Singha}
%\affiliation{Center for Astroparticle Physics and Space, Bose Institute,Block-EN,Sector-V, Salt lake , Bidhan Nagar , Kolkata- 700091, India} 
\author{Sabyasachi Ghosh}
\email{sabyaphys@gmail.com}
\affiliation { Department of Physics, University of Calcutta, 92, A.P.C. Road, Kolkata 700009, India}
%\affiliation{Department of Physics, Indian Institute of Technology, Mumbai, India 400076}
%\author{Guru Prakash Kadam}
%\email{guru@theory.tifr.res.in}
%%\affiliation{Theory Division, Physical Research Laboratory,
%Navrangpura, Ahmedabad 380 009, India}
%\affiliation{Department of Theoretical Physics, Tata Institute of Fundamental Research, Homi Bhabha Road, Mumbai 400005, India}
\author{Hiranmaya Mishra}
\email{hm@prl.res.in}
\affiliation{Theory Division, Physical Research Laboratory,
Navrangpura, Ahmedabad 380 009, India}
%\author{Amruta Mishra}
%\email{amruta@physics.iitd.ac.in}
%\affiliation{Department of Physics, Indian Institute of Technology, New 
%Delhi 110016, India}

\date{\today} 

\def\be{\begin{equation}}
\def\ee{\end{equation}}
\def\bearr{\begin{eqnarray}}
\def\eearr{\end{eqnarray}}
\def\zbf#1{{\bf {#1}}}
\def\bfm#1{\mbox{\boldmath $#1$}}
\def\hf{\frac{1}{2}}
\def\sl{\hspace{-0.15cm}/}
\def\omit#1{_{\!\rlap{$\scriptscriptstyle \backslash$}
{\scriptscriptstyle #1}}}
\def\vec#1{\mathchoice
        {\mbox{\boldmath $#1$}}
        {\mbox{\boldmath $#1$}}
        {\mbox{\boldmath $\scriptstyle #1$}}
        {\mbox{\boldmath $\scriptscriptstyle #1$}}
}

\begin{abstract}
We compute the transport coefficients, namely, the coefficients of shear and bulk viscosities as well as thermal
conductivity  for hot and dense matter. The calculations are performed 
within the Polyakov quark meson model.
The estimation of the transport coefficients is made using 
the Boltzmann kinetic equation within the relaxation time approximation. 
The energy-dependent relaxation time is estimated from meson meson 
scattering, quark meson scattering and quark quark scattering within the model.
In our calculations, the shear viscosity to entropy ratio and
the coefficient of thermal conductivity show a minimum at the critical
temperature, while the ratio of bulk viscosity to entropy density exhibits a
peak at this transition point.
 The effect of confinement modeled through a 
Polyakov loop potential plays an important role both below and above the critical temperature.

\end{abstract}


\maketitle

\section{Introduction}

 Transport coefficients of matter under extreme conditions of temperature, density or
external fields are interesting for several reasons. In the context of relativistic 
heavy ion collisions, these properties enter as dissipative coefficients in the 
hydrodynamic evolution of the quark gluon plasma that is produced following the collision ~\cite{heinzrev,Gyulassylarry2005,dirkprl,kapustalarryprl,luzumromatschke}.
Indeed, an extremely low value of the shear viscosity-to-entropy ratio ($\eta/s$) is needed to 
successfully describe the collective dynamics of the quark gluon matter at high temperature
and vanishing chemical potential to explain the elliptic flow data ~\cite{hirano,schenke}. 
At intermediate densities, near the chiral phase transition, which is being probed at the Facility for anti-proton and Ion Research(FAIR) program at Geselleschaft fuer Schwerionenforschung(GSI)-[8] and the Nuclotron-based Ion Collider fAcility(NICA) program at Joint Institute for Nuclear Research(JINR)-[9] motivates us to understand the behavior of transport coefficients at finite chemical potential and temperature.
motivates us to understand the behavior of transport coefficients at finite chemical potential and temperature.
Further, in the low temperature and high-density regime, the matter could be in one of the possible
types of color superconducting phases  of which transport properties also need to be understood ~\cite{rishi,pethick}. The cooling
of neutron stars at short time scales constrains the thermal conductivity ~\cite{chamel,sreddy} while the cooling through
neutrino emission on a much larger time scales constrains the phase of the matter in the interior of the compact star ~\cite{yakovlev,kaminker}. 
Further, the observable
regarding the viscosity of the such matter is the r-mode instability. In the absence of
viscous damping, the fluid in the rotating star becomes unstable to a mode that is coupled to gravity and radiates
away the angular momentum of the star ~\cite{rmode,jharmode}.
Apart from the wide variety of applications of the transport coefficients of strongly interacting matter, their temperature and chemical potential dependence may also be indicative of a phase transition
~\cite{kapustabk}. 

Transport coefficients for QCD matter in principle can be calculated using Kubo formulation ~\cite{kubo}. However, QCD is 
strongly interacting for both at energies accessible in heavy ion collision experiments as well as for the
densities expected to be there in the core of the neutron stars making the perturbative estimations unreliable.
Calculations using lattice QCD simulations at finite chemical potential is also challenging and is limited only
to the equilibrium thermodynamic properties at small chemical potentials.

The understanding of the elliptic flow in relativistic heavy ion collisions using hydrodynamics
with a low ($\eta/s$) and its connection to the conjectured lower bound ($\eta/s > 1/4\pi$)
using ADS/CFT correspondence ~\cite{kss} stimulated extensive investigation of this ratio for QCD matter.
These have been studied using perturbative QCD ~\cite{transqcd}, transport simulations of the  Boltzmann equation ~\cite{bassprl,phsdbrat},
 relaxation time approximation for solving the 
Boltzmann equations ~\cite{sasakinjl,klevansky,deb,voskresenskynpaa,voskresensky} and lattice simulation of QCD ~\cite{latticemeyer}.
Most of these calculations have been  performed at vanishing baryon density. The general variation of this ratio with
 temperature in most of these studies shows a minimum at the transition
temperature. The numerical value of $\eta$  at the minimum, however, differs by orders of magnitude. For example, 
Ref. \cite{itakura,lang}, Refs. \cite{nicola,mitra,prakash} have predicted $\eta$ of order 0.001 GeV$^3$,
$\eta$=0.002-0.003 GeV$^3$ while Ref. \cite{dobadoestrada} predicts a value of $\eta$ $\simeq$ 0.4 GeV$^3$. Further,
 the behavior of $\eta$/s shows a monotonic decrease with temperature in the Nambu-Jona-Lasinio (NJL) model in Ref. \cite{marty}.

 The bulk viscosity coefficient $\zeta$ has also been estimated in various effective models as well as in lattice QCD.
The rise of the bulk viscosity coefficient near the transition temperature has been observed in these effective models such as
chiral perturbation theory ~\cite{dobadoch}, quasiparticle models ~\cite{quasip}, linear sigma model ~\cite{purnendu},
and the Nambu-Jona-Lasinio model ~\cite{sasakinjl,deb}. 
Large bulk viscosity of matter produced in relativistic heavy ion collisions can give rise to different interesting phenomenon such as cavitation where pressure vanishes and hydrodynamic description of evolution becomes invalid 
 ~\cite{cavitation}. 
Here, again, the numerical value of the bulk viscosity coefficients
vary widely from $10^{-5}$ GeV$^3$ ~\cite{souravsuk} to $10^{-2}$ GeV$^3$ ~\cite{sasakinjl}.	

The other transport coefficient that is important at finite baryon density is the coefficient of
thermal conductivity $\lambda$ ~\cite{denicolhydro,denicolpre,denicolheat}. 
 The effects of thermal conductivity in relativistic hydrodynamics
has been discussed recently in Refs. ~\cite{rincon,denicolheat}. This coefficient has been evaluated in various
effective models like the Nambu-Jona-Lasinio model using the Green-Kubo approach \cite{fukutome},
relaxation time approximation ~\cite{deb}
and the  instanton liquid model ~\cite{nam}. The results, however,
 vary over a  wide range of  values, with 
$\lambda=0.008$ GeV$^{-2}$ as in Ref. ~\cite{nicola} 
to $\lambda \sim 10$ GeV$^{-2}$ as in Ref. ~\cite{marty} for a 
range of temperatures (0.12 GeV $<$T$<$ 0.17 GeV),  which  has 
been nicely tabulated in 
Ref. ~\cite{sabyath}. 

We shall attempt here to estimate these transport coefficients within
an effective model of strong interaction, the
Polyakov loop extended quark meson (PQM) model.
It has become quite popular during last few years due to its close relationship with the linear sigma model that captures the chiral symmetry breaking aspect
while being in agreement with the lattice QCD results for
 thermodynamics at vanishing baryon density. The physics of
 confinement is taken care of at least  partially 
 by coupling the quark field to the Polyakov loops so that 
quark excitations are suppressed below the transition temperature. Let us
 note that the transport coefficients like bulk viscosity apart from 
the distribution functions also depend upon the bulk thermodynamic 
quantities like velocity of sound. We wish to explore the effects of such nonperturbative properties on the 
transport coefficients.

 The transport coefficients are evaluated within the
relaxation time approximation of Boltzmann equation. 
The relaxation time is calculated by evaluating the scattering rates of the particles in the model, namely, 
the quarks and pion and sigma mesons, with their respective medium-dependent masses. The scattering processes 
considered here are meson scatterings as considered in Ref. ~\cite{purnendu},
 quark scattering through meson exchanges as in Refs.  ~\cite{sasakinjl,deb,marty},
and quark-meson scatterings.
As we shall see in the following, each of these processes brings out distinct
features for the transport coefficients.
We would like to mention here that these coefficients have also been 
estimated using Kubo formulation through 
one-loop self-energies for quarks and mesons  in a separate work ~\cite{pracheta}.

We organize the present investigation as follows. In the following section, we discuss the two-flavor PQM model
thermodynamics. 
The reason is that the expressions for transport coefficients involve meson masses which are medium dependent. Further, some transport coefficients like the bulk viscosity involves bulk thermodynamical properties such as energy density, pressure and the velocity of sound.
As the order parameters for chiral and confinement-deconfinement transitions are coupled, this leads to nontrivial relations for derivatives of the thermodynamic
potential with respect to external parameters like chemical potential or temperature as the mean fields themselves
are also medium-dependent. Furthermore, the implicit dependence of these mean fields/ order parameters are
calculated here analytically to avoid possible numerical errors. In Sec. III, we give the expressions
for the transport coefficients in terms of relaxation time and estimate them to finally give the results for these coefficients. We also compare them with the same obtained with alternate approaches
like the NJL model so that the effects of confinement-deconfinement transition
modeled through Polyakov loop potential is explicitly seen. Finally, we summarize and draw the conclusions of the present investigation in section IV.

\section{Thermodynamics of PQM model} 
We shall adopt here an effective model that captures two important features of QCD, namely, chiral symmetry 
breaking and its restoration at high temperature and/densities 
as well as the confinement-deconfinement transitions. Two such effective models
have become popular recently-- the Polyakov loop extended Nambu- Jona-Lasinio (PNJL) model and the Polyakov loop
extended  quark meson coupling model (PQM). These models are extensions 
respectively of NJL model and linear sigma model that captures various aspects of
chiral symmetry breaking pattern of strong interaction physics. Explicitly, the Lagrangian of the PQM model is given by ~\cite{bjschaefer,guptatiwari,bielich,buballa,ranjita}
\bearr
{\cal L}&=&\bar\psi\left(i\gamma^\mu D_\mu-m-g_\sigma(\sigma+i\gamma_5\bfm\tau\cdot\bfm\pi)\right)\psi\nonumber 
+ \frac{1}{2}\left[\partial_\mu\sigma\partial^\mu\sigma+\partial_\mu\bfm\pi\partial^\mu\bfm\pi\right]\nonumber - U_\chi(\sigma,\bfm\pi)-U_P(\phi,\bar\phi)\\
\label{lagpqm}
\eearr
In the above, the first term is the kinetic and interaction term for the quark doublet $\psi=(u,d)$ interacting with the scalar ($\sigma$)
and the isovector pseudoscalar pion $({\bfm \pi})$ field. The scalar field $\sigma$ and the pion field ${\bfm\pi}$ together form a SU(2)  isovector field.
 The quark field is also coupled to a spatially constant temporal gauge field $A_0$ 
through the covariant derivative $D_\mu=\partial_\mu-ieA_\mu$; $A_\mu=\delta_{\mu 0}A_\mu$.

The mesonic potential $U_\chi(\sigma,\zbf \pi)$ essentially describes 
the chiral symmetry breaking pattern in strong interaction and is given by
\be
U_\chi(\sigma,\bfm \pi)=\frac{\lambda}{4}(\sigma^2+\bfm \pi^2-v^2)^2-c\sigma
\label{uchi}
\ee

The last term in the Lagrangian in Eq. (\ref{lagpqm}) is responsible for including the physics of color confinement
in terms of a potential energy for the expectation value of the 
Polyakov loop $\phi$ and $\bar\phi$ which are defined in terms 
of the Polyakov loop operator which is a Wilson loop in the temporal direction
\be
{\cal P}=P\exp\left( i\int_0^\beta dx_0 A_0(x_0,\zbf x)\right).
\ee
In the Polyakov gauge  $A_0$ is time independent and is in the Cartan subalgebra i.e. $A_0^a=A_0^3\lambda_3+A_0^8\lambda_8$.
One can perform the integration over the time variable trivially as path ordering 
becomes irrelevant so that ${\cal P}(\zbf x)=\exp(\beta A_0)$.
The Polyakov loop variable $\phi$ and its Hermitian conjugate $\bar\phi$ 
are defined as
\be
\phi(\zbf x)=\frac{1}{N_c}Tr {\cal P(\zbf x)\quad\quad \bar\phi(\zbf x)}=\frac{1}{N_c}{\cal P}^\dagger(\zbf x).
\ee
In the limit of heavy quark mass, the confining phase is center symmetric and therefore $\langle\phi\rangle=0$, while
 for deconfined phase 
$\langle \phi\rangle\neq 0$. Finite quark masses break this symmetry explicitly.
 The explicit form of the potential $U_p(\phi,\bar\phi)$ is not known 
from first principle calculations. The common strategy is to choose a functional form of the potential that reproduces the pure gauge lattice
simulation thermodynamic results. Several forms of this potential has been suggested in literature. We shall use here the
polynomial parametrization ~\cite{bjschaefer}
\be
U_P(\phi,\bar\phi)=T^4\left[-\frac{b_2(T)}{2}\bar\phi\phi-\frac{b_3}{2}(\phi^3+\bar\phi^3)+\frac{b_4}{4}(\bar\phi\phi)^2\right]
\label{uphi}
\ee
with the temperature-dependent coefficient $b_2$ given as
\be
b_2(T)=a_0+a_1(\frac{T_0}{T})+a_2(\frac{T_0}{T})^2+a_3(\frac{T_0}{T})^3
\ee
The numerical values of the parameters are
\bearr
&&a_0=6.75,\quad a_1=-1.95, \quad a_2=2.625, \quad a_3=-7.44\nonumber\\
&&b_3=0.75, \quad b_4=7.5
\\
\label{parameters}
\eearr
The parameter $T_0$ corresponds to the transition temperature of Yang-Mills theory. However, for the full
dynamical QCD, there is a flavor dependence on $T_0(N_f)$. For two flavors we take it to be $T_0(2)=192$ MeV as in
Ref.~\cite{bjschaefer}.

The Lagrangian in Eq. (\ref{lagpqm}) is invariant under  $SU(2)_L\times SU(2)_R$ transformation when the
explicit symmetry breaking term $c\sigma$ vanishes in the potential $U_\chi$ in Eq. (\ref{uchi}).
 The parameters of the potential $U_\chi$ are chosen such that the chiral symmetry is spontaneously broken in the vacuum.
 The expectation values of the meson fields in vacuum are $\langle\sigma\rangle=f_\pi$ and $\langle\bfm\pi\rangle=0$. Here $f_\pi=93$ MeV 
is the pion decay constant. The coefficient of the symmetry breaking linear term is decided from the partial conservation of
 axial vector current as 
$c=f_\pi m_\pi^2$, $m_\pi=138$ MeV, being the pion mass. Then minimizing the potential one has $v^2=f_\pi^2-m_\pi^2/\lambda$.
 The quartic coupling for the meson,
$\lambda$ is determined from the mass of the sigma meson given as $m_\sigma^2=m_\pi^2+2\lambda f_\pi^2$. In the present work we take $m_\sigma=600 $MeV which gives $\lambda$=19.7. The coupling $g_\sigma$ is fixed here from the constituent quark mass in vacuum $M_q=g_q f_\pi$
 which has to be about one-third of the nucleon mass that leads to $g_\sigma=3.3$ ~\cite{rischkeqm}.

To calculate the bulk thermodynamical properties of the system we use 
a mean field approximation for the meson and the Polyakov fields while retaining the quantum and thermal fluctuations of the quark fields. The thermodynamic potential can then be written as
\be
\Omega(T,\mu)=\Omega_{\bar q q}+U_\chi+U_P(\phi,\bar\phi)
\label{thpot}
\ee
The fermionic part of the thermodynamic potential is given as
\be
\Omega_{\bar q q}=-2N_fT\int \frac{d^3 p}{(2 \pi)^3} 
\left[\ln\left(1+3(\phi+\bar\phi e^{-\beta\omega_-})e^{-\beta\omega_-}+e^{-3\beta\omega_-}\right)\\
+\ln\left(1+3(\phi+\bar\phi e^{-\beta\omega_+})e^{-\beta\omega_+}+e^{-3\beta\omega_+}\right)\right]
\ee
modulo a divergent vacuum part. In the above, $\omega_\mp=E_p\mp\mu$, with the single particle quark/antiquark energy $E_p=\sqrt{\zbf p^2+M^2}$.
The constituent quark/antiquark mass is defined to be
\be
M^2=g_\sigma^2(\sigma^2+\zbf \pi^2).
\ee
The divergent vacuum part arises from the negative energy states of the Dirac sea. Using standard renormalization,
it can be partly absorbed in the coupling $\lambda$ and $v^2$. However, a logarithmic correction from the renormalization
scale remains, and we neglect it  in the calculations that follow ~\cite{rischkeqm}. 

The mean fields are obtained by minimizing $\Omega$ with respect to $\sigma$, $\phi$, $\bar\phi$, and $\pi$. Extremizing
 the effective potential with respect to the $\sigma$ field leads to
\be
\lambda(\sigma^2+\zbf\pi^2-v^2)-c+g_\sigma\rho_s=0
\label{gapsigma}
\ee
where, the scalar density $\rho_s=-\langle\bar\psi\psi\rangle$ is given by
\be
\rho_s=6N_fg_\sigma\sigma\int\frac{d\zbf p}{(2\pi)^3}\frac{1}{E_P}\left [f_-(\zbf p)+f_+(\zbf p)\right].
\label{rhos}
\ee
In the above, $f_\mp(\zbf p)$ are the distribution functions for the quarks and anti quarks given as
\be
f_-(\zbf p)=\frac{\phi e^{-\beta\omega_-}+2\bar\phi e^{-2\beta\omega_-} 
+ e^{-3\beta\omega_-}}
{1+3\phi e^{-\beta\omega_-}+3\bar\phi e^{-2\beta\omega_-} + e^{-3\beta\omega_-}},
\ee
and,
\be
f_+(\zbf p)=\frac{\bar\phi e^{-\beta\omega_+}+2\phi e^{-2\beta\omega_+} 
+ e^{-3\beta\omega_+}}{1+3\bar\phi e^{-\beta\omega_+}+3\phi e^{-2\beta\omega_+} + e^{-3\beta\omega_+}},
\ee

The condition  $\frac{\partial\Omega}{\partial\phi}=0$ leads to
\be
T^4\left[-\frac{b_2}{2}\bar\phi-\frac{b_3}{2}\phi^2+\frac{b_4}{2}\bar\phi\phi\bar\phi\right]+I_\phi=0
\label{gapphi}
\ee
where ,
\be
I_\phi=\frac{\partial\Omega_{\bar q q}}{\partial \phi}=-6N_fT\int\frac{d\zbf p}{(2\pi)^3}
\left [\frac{e^{-\beta\omega_-}}{1+3\phi e^{-\beta\omega_-}+3\bar\phi e^{-2\beta\omega_-} + e^{-3\beta\omega_-}}
+\frac{e^{-2\beta\omega_+}}{1+3\bar\phi e^{-\beta\omega_+}+3\phi e^{-2\beta\omega_+} + e^{-3\beta\omega_+}}\right]
,
\ee
Similarly, $\frac{\partial\Omega}{\partial\bar\phi}=0$ leads to
\be
T^4\left[-\frac{b_2}{2}\phi-\frac{b_3}{2}\bar\phi^2+\frac{b_4}{2}\bar\phi\phi^2\right]+I_{\bar\phi}=0
\label{gapphib}
\ee
with,
\be
I_{\bar\phi}=\frac{\partial\Omega_{\bar q q}}{\partial \bar\phi}=-6N_fT\int\frac{d\zbf p}{(2\pi)^3}
\left[\frac {e^{-2\beta\omega_-}}{1+3\phi e^{-\beta\omega_-}+3\bar\phi e^{-2\beta\omega_-} + e^{-3\beta\omega_-}}
+\frac {e^{-\beta\omega_+}}{1+3\phi e^{-\beta\omega_+}+3\bar\phi e^{-2\beta\omega_+} + e^{-3\beta\omega_+}}
\right],
\ee

 Finally, minimization of the effective potential with respect to $\zbf \pi$ fields  leads to
\be
\frac{\partial\Omega}{\partial\zbf\pi}=\lambda(\sigma^2+\zbf\pi^2-v^2)\zbf\pi+g\zbf \rho_{ps}=0
\label{gappi}
\ee
where, the pseudoscalar density can be expressed as
\be
\zbf\rho_{ps}=\langle\bar q\i\gamma_5\zbf\tau q\rangle=6N_fg_\sigma\zbf\pi
\int\frac{d\zbf p}{(2\pi)^3}\frac{1}{E_P}\left [f_-(\zbf p)+f_+(\zbf p)\right].
\label{rhops}
\ee

The $\sigma$ and $\pi$ masses are  determined by the curvature of $\Omega$ at the global minimum
\be
M_\sigma^2=\frac{\partial^2\Omega}{\partial\sigma^2},\quad\quad M_{\pi_i}^2=\frac{\partial^2\Omega}{\partial\pi_i^2}.
\ee
These equations lead to  the masses for the $\sigma$ and pions given as
\be
M_\sigma^2=m_\pi^2+\lambda(3\sigma^2-f_\pi^2)+g_\sigma^2\frac{\partial\rho_s}{\partial\sigma}
\label{sigmass}
\ee
\be
M_\pi^2=m_\pi^2+\lambda(\sigma^2-f_\pi^2)+g_\sigma^2\frac{\partial\rho_{ps}}{\partial\pi}.
\label{pimass}
\ee

 Explicitly, using Eq. (\ref{rhos}),
\be
\frac{\partial\rho_s}{\partial\sigma}=\frac{6}{\pi^2}\int dp p^2\left[
\frac{g_\sigma p^2}{E(\zbf p)^3}\left(f_-(\zbf p)+f_+(\zbf p)\right)
+\frac{M}{E(\zbf p)}\left(\frac{\partial f_-}{\partial\sigma}+\frac{\partial f_+}{\partial\sigma}\right)\right]
\ee
The derivatives of the distribution functions with respect to to the scalar field $\sigma$ are given as
\be
\frac{\partial f_- (\zbf p)}{\partial \sigma}=
\frac{\beta g_\sigma^2 \sigma }{E(\zbf p)}
\left[3f_-^2-\frac{3 e^{-3 \beta \omega_- } +4\bar\phi e^{-2 \beta \omega_- }
+\phi e^{-\beta \omega_-}}{1+3\phi e^{-\beta\omega_-}+3\bar\phi e^{-2\beta\omega_-} + e^{-3\beta\omega_-}}
\right]
\ee

and,
\be
\frac{\partial f_+}{\partial \sigma}=
\frac{\beta g_\sigma^2 \sigma }{E(\zbf p)}
\left[3f_+^2-\frac{3 e^{-3 \beta \omega_+ } +4\phi e^{-2 \beta \omega_+ }
+\bar\phi e^{-\beta \omega_+}}{1+3\bar\phi e^{-\beta\omega_+}+
3\phi e^{-2\beta\omega_+} + e^{-3\beta\omega_+}}
\right]
\ee

Similarly, using Eq. (\ref{rhops})
\be
\frac{\partial\rho_{ps}}{\partial \pi}=
\frac{6}{\pi^2}\int dp \frac{p^2}{E(\zbf p)}\left[f_-(\zbf p)+f_+(\zbf p)\right].
\ee
In the above we have set the expectation value of pion field to be zero, i.e. $\bfm\pi=0$ so that constituent quark mass is $M^2=g_\sigma^2\sigma^2$.

 The net quark density is given by,
\be
n=-\frac{\partial\Omega}{\partial\mu}=\frac{6}{\pi^2}\int p^2dp\left[f_-(\zbf p)-f_+(\zbf p)\right]
\label{rhob}
\ee

\begin{figure}[t!]
\vspace{-0.4cm}
\begin{center}
\begin{tabular}{c c}
\includegraphics[width=9cm,height=9cm]{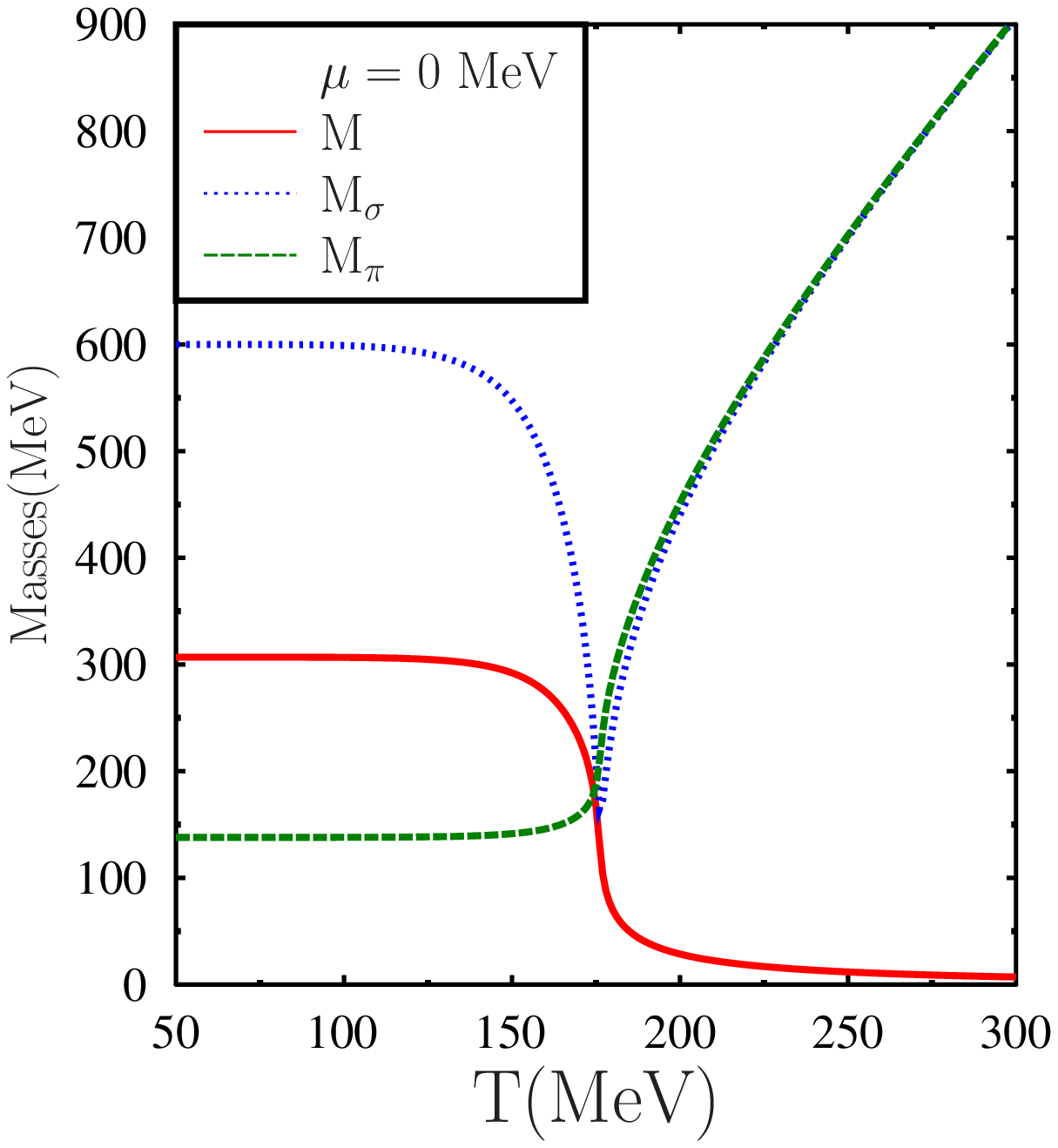}&
\includegraphics[width=9cm,height=9cm]{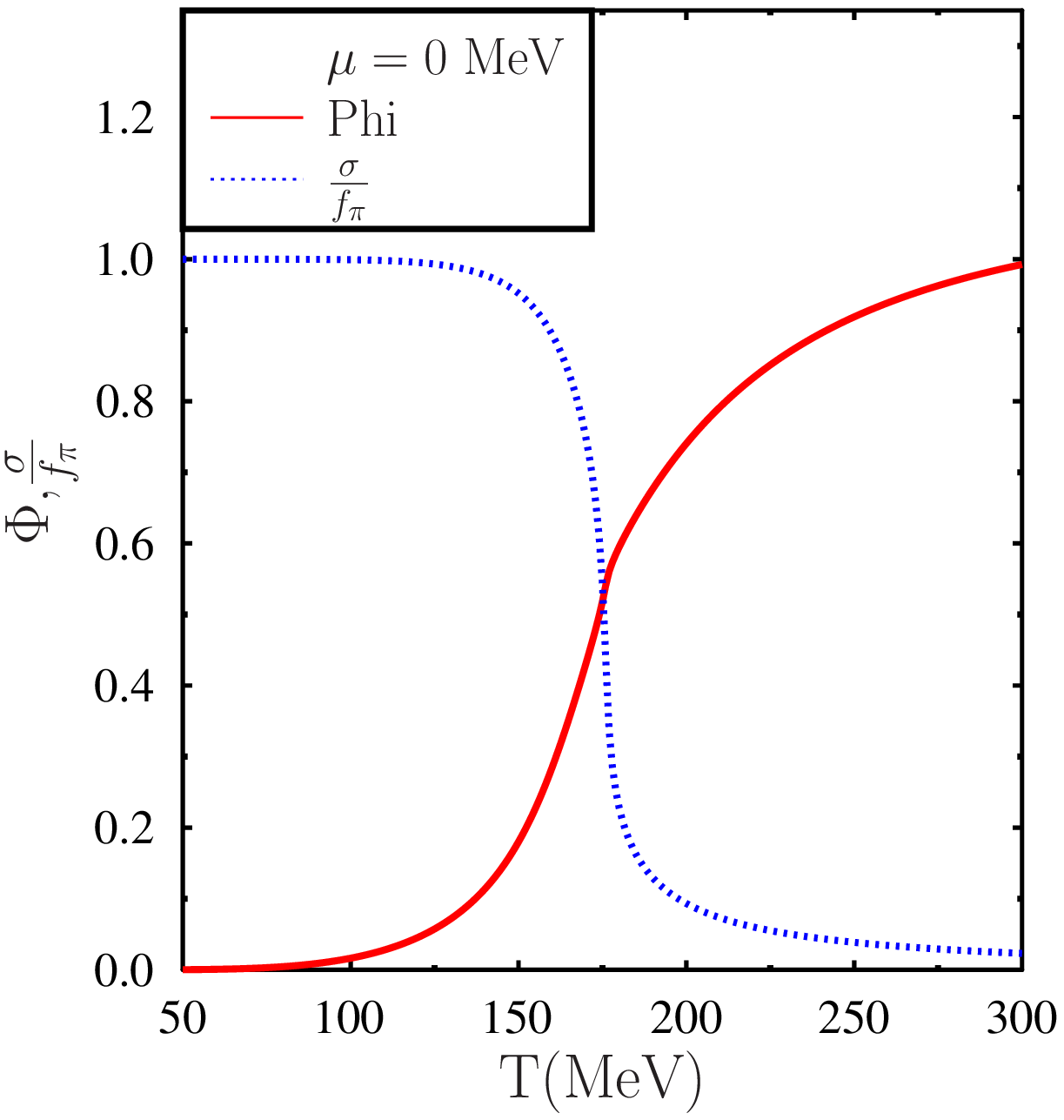}\\
(a) & (b)
\end{tabular}
\end{center}
\caption{ (a) Temperature dependence of the masses of constituent quarks ($M$), and pions ($M_\pi$) and sigma mesons ($M_\sigma$)
 and ~(b)
 the order parameters $\sigma$ and $\phi$ as a function of temperature for  $\mu=0$ MeV .}
\label{fig1}
\end{figure}

The energy density $\epsilon=\Omega-T\partial\Omega/\partial T+\mu\rho_q$ is given by
\be
\epsilon=\frac{6}{\pi^2}\int p^2dp E(\zbf p)\left(f_-(\zbf p)+f_+(\zbf p)\right)+U_\chi-3 U_P(\phi,\bar\phi)
+\frac{T^5}{2}\frac{db_2(T)}{dT}\bar\phi \phi
\label{energy}
\ee
In Fig.~\ref{fig1}(a), we have plotted the constituent quark mass, and the meson masses as given in Eq. (\ref{sigmass}) and Eq. (\ref{pimass}) 
as a function of temperature for vanishing baryon density. In the chirally broken phase, the pion mass, being the mass
of an approximate Goldstone mode is protected and varies weakly with temperature. On the other hand,
the mass of $\sigma$ , $M_\sigma$,
which is approximately twice the constituent quark mass,$M$ drops significantly near the
crossover temperature. At high temperature, being chiral partners, the masses of $\sigma$ and $\pi$ mesons become degenerate and increase linearly with temperature.
 In Fig.~\ref{fig1}(b), we have plotted the order parameters $\sigma$ and $\phi$ as a function of temperature for vanishing quark chemical potential.
 We also note that for $\mu=0$, the order parameters $\phi$ and $\bar\phi$ are the same. 
Because of the approximate chiral symmetry, the chiral order parameter  
decreases  with temperatures to  small values but never vanishes. The Polyakov loop parameter on the other hand 
grows from $\phi=0$ at zero temperature to about $\phi=1$ at high temperatures. We might mention here that at very high temperature the value of polyakov loop parameter exceeds unity, the value in the infinite quark mass limit.

%\newpage
\hfill
\vfill
\begin{figure}
\vspace{-0.4cm}
\begin{center}
\includegraphics[width=9cm,height=9cm]{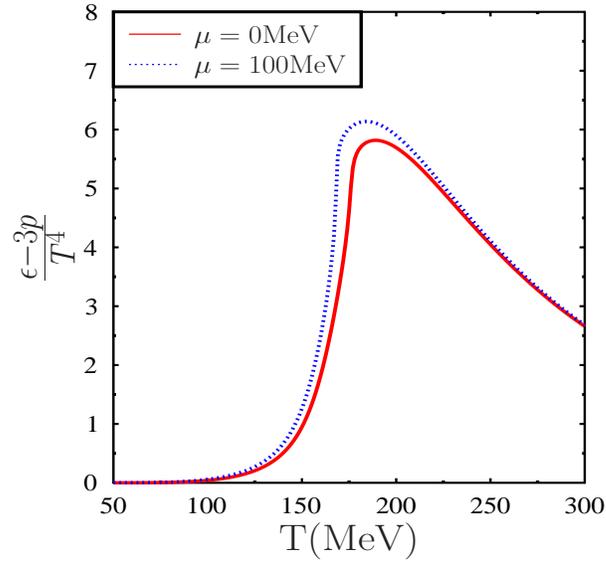}
\end{center}
\caption{  Temperature dependence of the scaled trace anomaly $\tfrac{\epsilon-3p}{T^4}$
}
\label{fig2}
\end{figure}

\begin{figure}
\vspace{-0.4cm}
\begin{center}
\begin{tabular}{c c}
\includegraphics[width=9cm,height=9cm]{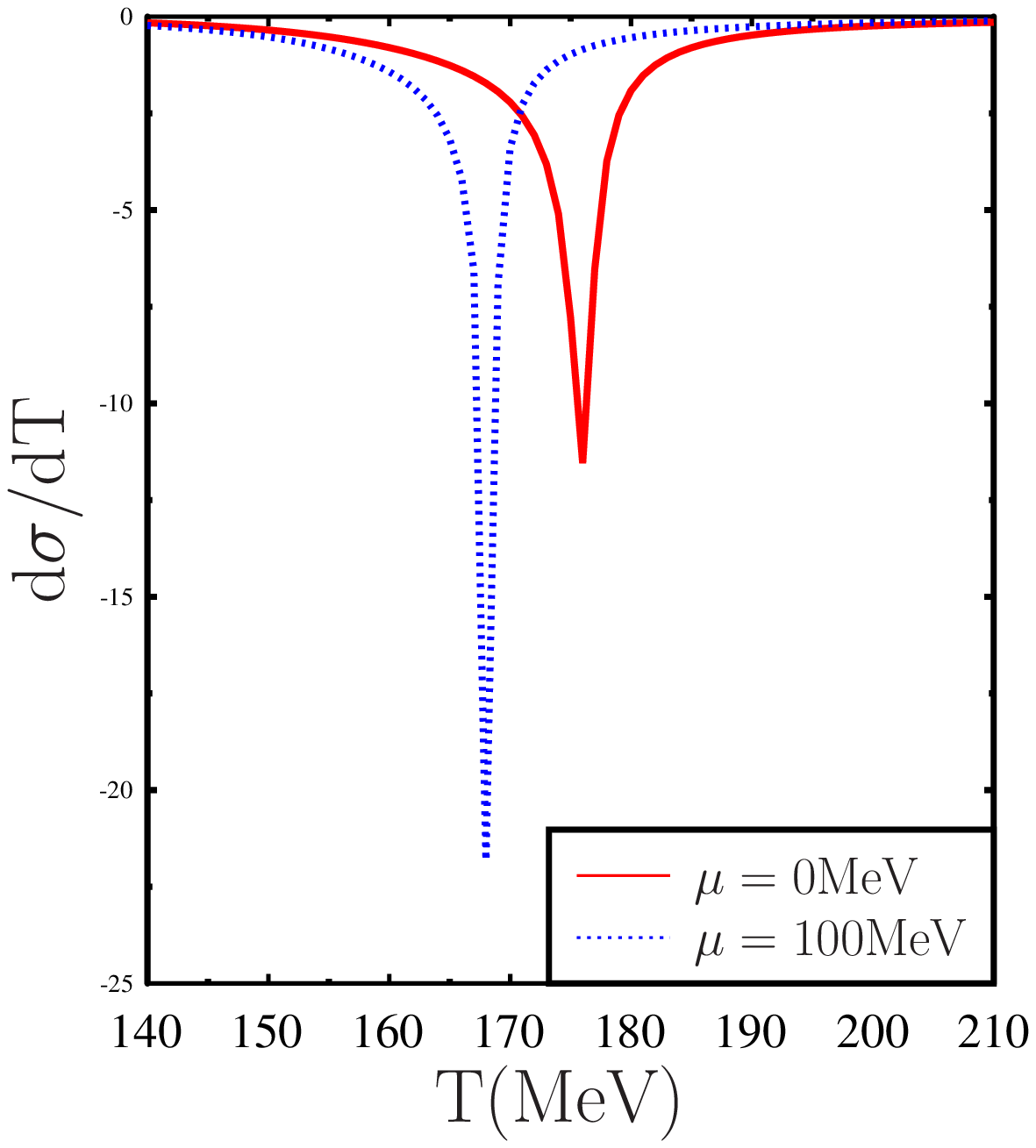}&
\includegraphics[width=9cm,height=9cm]{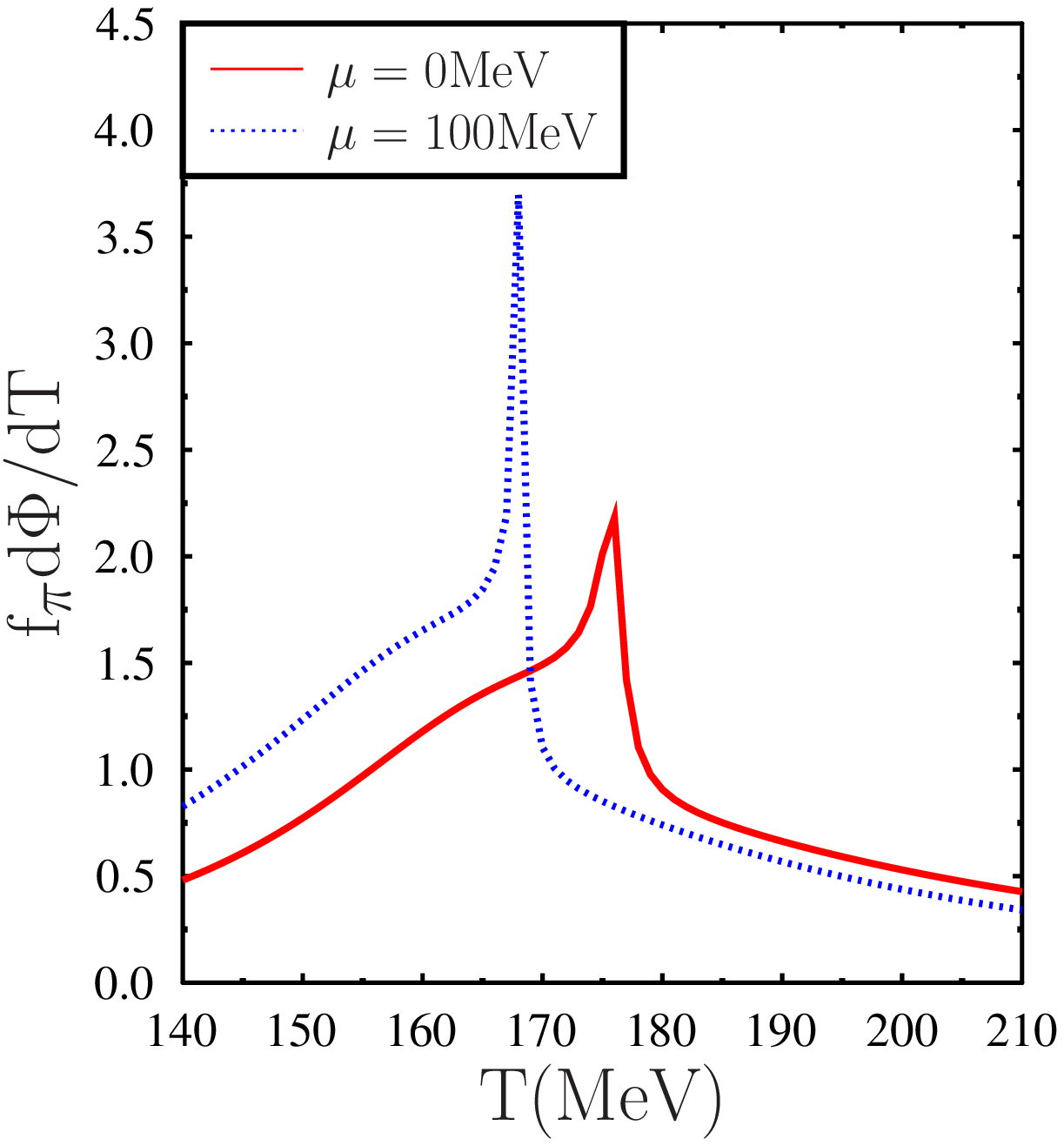}\\
(a) & (b)
\end{tabular}
\end{center}
\caption{ (a) Temperature derivative of the chiral order parameter $(\tfrac{d\sigma}{dT})$ and (b) Polyakov loop parameter
 $(\tfrac{d\phi}{dT})$
as a function of temperature .}
\label{fig3}
\end{figure}

 Next, in Fig.~\ref{fig2}, 
we show the dependence of the 
trace anomaly $(\epsilon-3p)/T^4$ on temperature. The conformal symmetry is broken maximally 
at the critical temperature.  Further finite chemical potential
enhances this breaking as it breaks scale symmetry explicitly. As we shall see later, this will have its implication on
the bulk viscosity coefficients.

 %In Fig 3-b, we show the 
 %the temperature dependence of the square of the  velocity of sound $v_n^2=(dp/d\epsilon)_n$ at constant 
%quark number density as defined in Eq.\ref{sound}.

%\newpage
%\section{Analytic calculation of derivatives}
Next, to discuss critical behavior as well as to calculate different thermodynamic quantities one has to take derivatives
 of the thermodynamic potential with respect to the mean fields as well as the parameters like temperature and the
chemical potential.  Vanishing of  the first order derivatives of the thermodynamic potential  with respect
to the order parameters leads to the values 
of the order parameters satisfying the coupled gap equations as shown. However, to calculate many different 
thermodynamic quantities one also has to take into account the implicit dependence of the order 
parameters on temperature as well as chemical potential. One can do a numerical differentiation of the 
order parameters after solving for them from the gap equation. However, this can be numerically less accurate particularly 
for the higher-order derivatives. We shall use here a semianalytic  approach to calculate the implicit contributions to the extent of taking the
differentiation of the expressions analytically ~\cite{ghoshraha}.
Only the values of the final expressions so obtained are
computed numerically. For example, to calculate the derivative of the order parameter $X $,
 $(X=\sigma,\phi,\bar\phi)$ with respect
 to temperature is given by the equation
\be
\frac{\partial}{\partial T}\left(\frac{\partial\Omega}{\partial X}\right) 
+\frac{\partial}{\partial \sigma}\left(\frac{\partial\Omega}{ \partial X}\right) \frac{d\sigma}{d T}
+\frac{\partial}{\partial \phi}\left(\frac{\partial\Omega}{ \partial X}\right) \frac{d\phi}{d T}
+\frac{\partial}{\partial \bar\phi}\left(\frac{\partial\Omega}{ \partial X}\right) \frac{d\bar\phi}{d T}=0.
\label{opderivative}
\ee
Thus we have a matrix equation of the type $\bf A\cdot \bf Y=\bf B$, where $\bf A$ is the coefficient matrix of the variables 
$\bf Y=\left(\tfrac{d\sigma}{dT},\tfrac{d\phi}{dT},\tfrac{d\bar\phi}{dT}\right)^T$, and $\bf B$ is the matrix of 
derivatives of the thermodynamic potential involving explicit dependence on temperature, i.e.,
 $\bf B=\left(-\tfrac{\partial}{\partial T}(-\tfrac{\partial \Omega}{\partial\sigma},-\tfrac{\partial \Omega}{\partial\phi},
-\tfrac{\partial \Omega}{\partial\bar\phi})^T\right)$.These
 matrix equations can be solved using Cramers rule. The coefficient matrix $\bf A$ is given by

\begin{equation}
\bf A =
\left[
\begin{array}{ccc}
\Omega_{\sigma\sigma} &\Omega_{\sigma\phi}&\Omega_{\sigma\bar\phi}\\
\Omega_{\phi\sigma} &\Omega_{\phi\phi}&\Omega_{\phi\bar\phi}\\
\Omega_{\bar\phi\sigma} &\Omega_{\bar\phi\phi}&\Omega_{\bar\phi\bar\phi}
\end{array}
\right]
\end{equation}
with, $\Omega_{ab}=\tfrac{\partial^2\Omega}{\partial a\partial b}$ where $a,b$ stand for $\sigma,\phi$ and $\bar\phi$.  Similarly, to calculate the derivatives with respect to chemical potential, 
the coefficient matrix $\bf A $ remains the same while the matrix $\bf B$ will involve derivatives of the thermodynamic potential involving
explicit dependence on the chemical potential.

Solving Eq. (\ref{opderivative}) this way, we have plotted the derivatives of the order parameters in Fig.~\ref{fig3}. The critical 
temperature is defined by the position of the peaks of these derivatives of the order parameters. At zero chemical potential this occurs at $T_C\simeq 176$ MeV.
Let us note that at $T_C$, the quark mass is $m_q=g_\sigma \sigma=134 $MeV, while the Polyakov loop variable $\phi\sim =0.5$. Thus at the critical temperature the effect of interaction is significant.
As chemical potential
for the quarks increase the critical temperature decreases. With finite chemical potential the peaks also become
sharper and at higher chemical potential the transition becomes a first order one. The critical point within this model 
occurs at $(T_c,\mu_c)=(155,163)$ MeV.

The other thermodynamic quantity that enters into the transport coefficient calculation is the velocity of sound. 
The same at constant density
is defined as
\be
c_s^2= \left(-\frac{\partial P}{\partial \epsilon}\right)_n=
\frac{s\chi_{\mu\mu}-\rho\chi_{\mu T}}{T(\chi_{TT}\chi_{\mu\mu}-\chi_{\mu T}^2)}
\label{sound}
\ee
where, 
$P$,the pressure, is the negative of the thermodynamic potential given in Eq. (\ref{thpot}). Further, 
$s=-\tfrac{\partial\Omega}{\partial T}$ is the entropy density and  the susceptibilities are defined as 
$\chi_{xy}=-\tfrac{\partial^2\Omega}{\partial x\partial y}$.
The velocity of sound shows a minimum near the crossover temperature. Within the model, at low temperature when the constituent quarks
start contributing to the pressure, their contribution to the energy density is significant compared to their contribution to the pressure
leading to decreasing behavior of the velocity of sound until the crossover temperature, beyond which it increases as the quarks become 
light and approach the massless limit of $c_s^2=\tfrac{1}{3}$. 
Such a dip in the velocity of sound is also observed in lattice simulation ~\cite{tanmoy}.
 As we shall observe later, this behavior will have important consequences for the
behavior of bulk viscosity as a function of temperature.
We might mention here that such a dip for the sound velocity was not observed for two-flavor NJL ~\cite{deb}. For the linear sigma model calculations such a dip was observed only for a large sigma meson mass ~\cite{purnendu}.
\begin{figure}
\vspace{-0.4cm}
\begin{center}
\includegraphics[width=9cm,height=9cm]{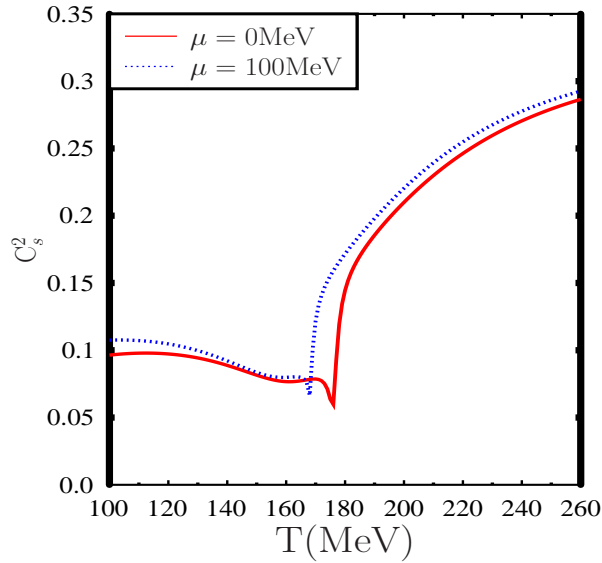}
\end{center}
\caption{  Temperature dependence of the velocity of sound at constant density. }
\label{fig4}
\end{figure}
\section{ Transport coefficients in relaxation time approximation}
We shall attempt here to estimate the transport coefficients in the relaxation time approximation where the particle masses are medium 
dependent. Such attempts were made earlier for the $\sigma$-model~ ~\cite{purnendu} as well as in the NJL model to
 compute the shear and bulk viscosity
coefficients. Such an approach was also made to estimate the viscosity coefficients of pure 
gluon matter ~\cite{blumredlich}. In all these attempts, the expressions for the viscosity coefficients
were derived for vanishing chemical potential. Several attempts were made to estimate these coefficients
 with finite chemical potential
with different Ansatze. These expressions were put on  firmer ground by deriving the expressions when there are 
mean fields and medium-dependent masses in a quasiparticle picture ~\cite{albright}. The resulting expressions for
 the transport coefficients were manifestly positive definite as they should be.
These expressions were derived explicitly for the NJL model ~\cite{deb}.
We use the same expressions here for the transport coefficients. The shear viscosity coefficient is given
by
\be
\eta=\frac{1}{15 T} \sum_a \int \frac{d\zbf p}{(2\pi)^3} \frac {p_a^4}{E_a^2}\tau(E_a)f_a^0(1\pm f_a^0)
\label{eta}
\ee
where, the sum is over all the different species contributing to the viscosity coefficients including the antiparticles,
and, $\tau^a$ is the energy-dependent relaxation time that we define in the following subsection.
The coefficient of bulk viscosity is given by
\bearr
\zeta &=&\frac{1}{9T} \sum_a\int \frac{d\zbf p}{(2\pi )^3}\frac{\tau ^a}{E_a{}^2}
f_a{}^0\left(1\pm f_a{}^0\right) \bigg[\zbf p^2\left(1-3v_n{}^2\right)-3v_n{}^2\left(M^2-TM \frac{dM}{dT}-{\mu M}
 \frac{{dM}}{d\mu }\right)\nonumber\\
&+&3\left(\frac{\partial P}{\partial n}\right)_{\epsilon }\left(M \frac{dM}{d\mu }-E_at^a\right)\bigg]^2
\label{zeta}
\eearr

The thermal conductivity on the other hand is given by
\be
\lambda=\left(\frac{w}{nT}\right)^2\sum_a\int \frac {d\zbf p}{(2\pi)^3}\frac{\zbf p^2}{3 E_a^2}\tau_a(E_a)
\left(t_a-\frac{nE_a}{w}\right)^2 f_a^0(1\pm f_a^0)
\label{lamb}
\ee

 In the above expressions, $f_a^0$ is the equilibrium  fermion/boson distribution functions depending upon the statistics
with $(1\pm f_a^0)$ being the Bose enhancement/ Fermi suppression factors and $t_a$ = +1, − 1 and 0 for the quark, antiquark, meson
respectively. Further,
 $c_s^2=\left(\frac{\partial p}{\partial \epsilon}\right)_n$ is the velocity of
sound at constant density and $w=\epsilon+p$ is the enthalpy density.
\subsection{Relaxation time estimation- meson scatterings}
As may be noted, the expressions for the  transport coefficients as in Eqs. (\ref{eta},\ref{zeta},\ref{lamb}),
depend not only on bulk thermodynamic properties like energy density, pressure, velocity of sound but also on the
energy-dependent relaxation time $\tau(E)$. In the following we shall first
estimate the relaxation times involving meson exchanges similar to Ref. ~\cite{purnendu}.

%\section*{ Scattering amplitudes in PQM Model}
Using the Lagrangian Eq. (\ref {lagpqm}), we calculate the relaxation time in PQM model by taking into account the following scattering amplitudes with the
corresponding matrix elements being given as  \\
%1. $\sigma  + \sigma  \to  \sigma  + \sigma$ \\
%2. $\sigma + \pi \to \sigma + \pi$ \\
%3. $\pi + \pi \to \pi + \pi $\\
%4. $\pi  + \pi  \to  \sigma  + \sigma$\\

%The corresponding amplitudes are :\\
\begin{equation}
M_{\sigma  + \sigma  \to  \sigma  + \sigma}=-6\lambda - 36 \lambda ^2f_{\pi }{}^2\left(\frac{1}{s-m_{\sigma }{}^2}+\frac{1}{t-m_{\pi }{}^2}+\frac{1}{u-m_{\pi }{}^2}\right)\\
\end{equation}

\begin{equation}
M_{\bfm \pi + \sigma \to \bfm \pi + \sigma}=-2\lambda - 4 \lambda ^2f_{\pi }{}^2\left(\frac{3}{t-m_{\sigma }{}^2}+\frac{1}{u-m_{\pi }{}^2}+\frac{1}{s-m_{\pi }{}^2}\right)\\
\end{equation}
\begin{equation}
M_{\bfm \pi + \bfm \pi \to \bfm \pi + \bfm \pi}=-2 \lambda \left(\frac{s-m_{\pi }{}^2}{s-m_{\sigma }{}^2}\delta _{\text{ab}}\delta _{\text{cd}}+ \frac{t-m_{\pi }{}^2}{t-m_{\sigma }{}^2}\delta _{\text{ac}}\delta _{\text{bd}}+\frac{u-m_{\pi }{}^2}{u-m_{\sigma }{}^2}\delta _{\text{ad}}\delta _{\text{bc}}\right) \\
\end{equation}
\begin{equation}
M_{\bfm \pi  + \bfm \pi  \to  \sigma  + \sigma}=-6\lambda - 4 \lambda ^2f_{\pi }{}^2\left(\frac{3}{s-m_{\sigma }{}^2}+\frac{1}{t-m_{\pi }{}^2}+\frac{1}{u-m_{\pi }{}^2}\right)\\
\end{equation}

 The terms involving the propagators yield divergent integrals due to the poles in s and u channel which is known in the 
literature ~\cite{purnendu}. To regulate these integrals one can include a width for the mesons as evaluated in the next subsection 
(Eq. (\ref{impi})). However, such a substitution violates crossing symmetry. Further,  these terms are generated from the three-point 
vertices which are not taken into 
account in the mean field approximation used in solving the gap equations and the resulting equation of state. Hence,
  to be consistent with equation of state while maintaining crossing symmetry for the scattering amplitudes,
we approximate the above scattering amplitudes by their limits when 
$s$, $t$ and $u$ are taken to be infinity and the scattering amplitudes reduce to constants ~\cite{purnendu}. Thus,
 the scattering amplitudes essentially reduce to constants. This allows us to compare our results with earlier work of ~\cite{purnendu}
 and study the effect of Polyakov loop and quarks within similar approximation.

The  energy-dependent interaction frequency $\omega_a(E_a)$ 
 for the particle specie $`a`$ arising from a scattering process 
$a,b\rightarrow c,d$, which is also the inverse of the energy-dependent relaxation time $\tau(E_a)$
is given by, with $d\Gamma_i=\frac{d\zbf p_i}{2 E_i(\zbf p)(2\pi)^3}$, ~\cite{deb}
\begin{equation}
\omega(E_a)\equiv\tau(E_a)^{-1}=\sum_b\int d \Gamma_b f_b^0W_{ab}(s).
\label{weboson}
\end{equation}
In the above, the summation is over all the particles  except the species $a$ with $a,b$ as the initial state.

The quantity $W_{ab}$ is dimensionless, Lorentz-invariant, and depends only on the Mandelstam variable $s$ and is given by

\bearr
W_{ab}(s) &=&\frac{1}{1+\delta_{ab}} \int d\Gamma_{c}d\Gamma_d (2\pi)^4\delta^4(p_a+p_b-p_c-p_d)\nonumber\\
&\times & |M|^2 (1+f_c)(1+f_d) 
\eearr
In the above, we have included the Bose enhancement factors for the meson scattering. The quantity 
$W_{ab}(s)$ is related to the cross section by noting that, with $t$ as the Mandelstam variable $t=(p_a-p_c)^2$,
\be
\frac{d\sigma}{dt}=\frac{1}{64\pi s}\frac{1}{p_{ab}^2} |M|^2
\ee
where,  $p_{ab}(s)=1/(2\sqrt{s})\sqrt{\lambda(s,m_a^2,m_b^2)}$, and  the kinematic function $\lambda (x,y,z)=x^2+y^2+z^2
-2 xy-2yz-2zx$, is the magnitude of the 3-momentum of the incoming particle in the c.m. frame. 
In the c.m. frame, using the energy momentum-conserving delta function and integrating over the final momenta, we have
\be
W_{ab}(s)=\frac{ 4\sqrt{s} p_{ab}(s)}{1+\delta_{ab}}\int_{t_{min}}^{t_{max}} dt \left(\frac{d\sigma}{dt}\right)
 (1+f_c(E_c))(1+f_d(E_d)).
\label{wabb}
\ee
where, 
$$t_{max,min}=m_a^2+m_c^2-\frac{1}{2s}(s+m_a^2-m_b^2)(s+m_c^2-m_d^2)\pm\frac{1}{2s}\sqrt{\lambda(s,m_a^2,m_b^2)\lambda(s,m_c^2,m_d^2)}$$
In the limit of constant $|M|^2$,  Eq. (\ref{wabb})  reduces to
\be
W_{ab}(s)=\frac{1}{1+\delta_{ab}}\frac{|M|^2}{16\pi\sqrt{s}p_{ab}}\left(t_{max}-t_{min}\right) (1+f_c(E_c))(1+f_d(E_d))
\ee
and, the transition frequency or the inverse relaxation time is given as
\be
\omega(E_a)\equiv \tau(E_a)^{-1}=\frac{1}{256\pi^3 E_a}\int_{m_b}^\infty dE_b \sqrt{E_b^2-m_b^2}f(E_b)|M|^2\int_{-1}^{1}
\frac{dx}{1+\delta_{ab}}\frac{1}{p_{ab}\sqrt{s}}\left(t_{max}-t_{min}\right).
\label{omegaea}
\ee
In the above,
$$s=2E_aE_b\left(1+\frac{m_a^2+m_b^2}{2E_aE_b}-\frac{p_ap_b}{E_aE_b}x\right)$$
To calculate e.g. the $\pi^+$ relaxation time ($\tau_{\pi^+}$), we consider the scattering 
processes $\pi^++\pi^i \to \pi^++\pi^i$ ($i=+,-,0$)
and, $\pi^++\sigma\to\pi^++\sigma$. 

To get an order of magnitude of the average relaxation time, one can also calculate an energy averaged 
 mean interaction frequency  for a given species as
$\bar\omega_a\equiv{\bar{\tau}_a}^{-1}$ as
\be
\bar\omega_a=\frac{1}{n_a}\int\frac{d\zbf p}{(2\pi)^3} \omega_a(E_a)f_a(E_a),
\label{taubarpi}
\ee
with 
\be
n_a=\int\frac{d\zbf p}{(2\pi)^3}f_a(E_a).
\ee

\subsection{Relaxation time estimation-- Quark scatterings}
We next consider the quark scattering within the model through the exchange of pion and sigma meson resonances. The approach is similar to 
Refs. ~\cite{deb,klevansky,marty} performed within NJL model to estimate the corresponding relaxation time for the quarks and
antiquarks.  The transition frequency is again given by Eq. (\ref{weboson}), with the corresponding $W_{ab}$ given as
\be
W_{ab}^q(s)=\frac{2\sqrt{s(s-4m^2)}}{1+\delta_{ab}}\int_{t_{min}}^{0}dt\left(\frac{d\sigma}{dt}\right)
\left(1-f_c(\frac{\sqrt s}{2},\mu)\right)\left(1-f_d(\frac{\sqrt s}{2},\mu)\right)
\label{wsig}
\ee
 where,
\be
\frac{d\sigma}{dt}=\frac{1}{16\pi s (s-4m^2)} \frac{1}{p^2_{ab}} |\bar M|^2
\ee
with the corresponding suppression factors appropriate for fermions.  For the quark scatterings, in the present case for 
two flavors we consider the following scattering processes:
$$u\bar u\rightarrow u\bar u,\quad u\bar d\rightarrow u\bar d,\quad u\bar u\rightarrow d\bar d,$$
$$u u\rightarrow u u,\quad u d\rightarrow u d,\quad \bar u\bar u\rightarrow\bar u \bar u,$$
$$\bar u\bar d\rightarrow \bar u\bar d,\quad d\bar d\rightarrow d\bar d,\quad d\bar d\rightarrow u\bar u,$$
$$d\bar u\rightarrow d\bar u,\quad d d\rightarrow d d,\quad \bar d\bar d\rightarrow \bar d\bar d,$$
One can use $i$-spin symmetry, charge conjugation symmetry and crossing symmetry to relate the matrix element
square for the above 12 processes to get them related to one another and one has to evaluate only two independent matrix elements
to evaluate all the 12 processes. We  choose these, as in Ref. ~\cite{klevansky}, to be the processes
$u\bar u\rightarrow u\bar u$ and $u\bar d\rightarrow u\bar d$ and use the symmetry conditions to calculate the rest. We note, however,
that, while the matrix elements are related, the thermal-averaged rates are not,  as they involve also the thermal
distribution functions for the initial states as well as the Pauli blocking factors for the final states.  
We also write down the square of the  matrix elements for these two processes explicitly ~\cite{deb,klevansky}--.
\bearr
|\bar M_{u\bar u\rightarrow u\bar u}|^2 &=&
g_\sigma^4\bigg [s^2|D_\pi(\sqrt s,0)|^2
+t^2|D_\pi(0,\sqrt{-t})|^2
(s-4m^2)^2|D_\sigma(\sqrt s,0)|^2
+(t-4m^2)^2|D_\sigma(0,\sqrt{-t})|^2\nonumber\\
&+&\frac{1}{N_c}Re\bigg(st D_\pi^*(\sqrt s,0)D_\pi(0,\sqrt{-t})
+s(4m^2-t)D_\pi^*(\sqrt s,0)D_\sigma(0,\sqrt{-t})\nonumber\\
&+&t(4m^2-s)D_\pi(0,\sqrt{-t})D_\sigma^*(\sqrt s,0)
+(4m^2-s)(4m^2-t)D_\sigma(0,\sqrt{-t})D_\sigma^*(\sqrt s,0)\bigg )\bigg].
\label{uubtouub}
\eearr
Similarly, the same for the process $u\bar d\rightarrow u\bar d$ is given as ~\cite{klevansky}
\bearr
|\bar M_{u\bar d\rightarrow u\bar d}|^2 &=&
 g_\sigma^4\bigg[4s^2|D_\pi(\sqrt s,0)|^2
+t^2|D_\pi(0,\sqrt{-t})|^2
(s-4m^2)^2|D_\sigma(\sqrt s,0)|^2
+(t-4m^2)^2|D_\sigma(0,\sqrt{-t})|^2\nonumber\\
&+&\frac{1}{N_c}Re\bigg(-2st D_\pi^*(\sqrt s,0)D_\pi(0,\sqrt{-t})
+2s(4m^2-t)D_\pi^*(\sqrt s,0)D_\sigma(0,\sqrt{-t})\bigg)\bigg].
\label{udbtoudb}
\eearr
The meson propagators $D_{a}(\sqrt{s},0)$, ($a=\sigma,\zbf \pi$) is given by
\be
D_{a}(\sqrt{s},\zbf 0)=\frac{i}{s-M_a^2-i Im\Pi_{M_a}(\sqrt{s},\zbf 0)}
\label{propmeson}
\ee
In the above, the masses of the mesons are given by Eqs. (\ref{sigmass}) and (\ref{pimass}) determined by the curvature of 
the thermodynamic potential.
Further, in Eq. (\ref{propmeson}), $Im\Pi(\sqrt{s},0)$ which is related to the width of the resonance as 
$\Gamma_a=Im\Pi_a/M_a$ is given as ~\cite{klevansky}
\be
Im\Pi_a(\omega,\zbf 0)=\theta(\omega^2-4 m^2)\frac{N_cN_f}{8\pi\omega}\left(\omega^2-\e_a^2\right)\sqrt{\omega^2-4m^2}
\left(1-f_-(\omega)-f_+(\omega)\right)
\label{impi}
\ee
with $\e_a=0$ for pions and $\e_a=2m$ for sigma mesons.

With the squared matrix elements for the quark scatterings given as above the transition frequency for the quark of a given species
is
\be
\omega_q(E_a)=\frac{1}{2E_a}\int d\pi_b f(E_b) W_{ab}^q.
\label{omgeaq}
\ee
\subsection{Quark pion scattering and relaxation time}

Next, we compute the contribution of quark meson scattering to the relaxation times for both mesons as well as
quarks. One can argue that the dominant contribution comes from pions as  their number is large
 compared to the sigma mesons
both below and above $T_c$. Therefore, in the following we consider the quark-pion scattering only. The Lorentz-invariant scattering matrix element
can be written as $\bar U(p_2)T_{ba}U(p_1)$, with $\bar UU=2m_q$ and with $p_1,p_2$ denoting the
initial and the final  quark momenta, respectively, and $q_1,q_2$, being the momenta of the pions.
\be
T_{ba}=\delta_{ba}\frac{1}{2}(q_1+q_2)^\mu\gamma_\mu (\delta_{ab}B^{(+)}+i\epsilon_{abc}\tau_c B^{(-)})
\ee

 where,
\be
B^{(+)}=g_\sigma^2\left(\frac{1}{u-m_q^2}-\frac{1}{s-m_q^2}\right),
\label{bplus}
\ee
and
\be
B^{(-)}=-g_\sigma^2\left(\frac{1}{u-m_q^2}+\frac{1}{s-m_q^2}\right).
\label{bminus}
\ee

Averaging over the spin and isospin factors, the matrix element square for the quark-pion scattering is given by
\be
|\bar M|^2=\frac{g_\sigma ^4}{6}\left((s-u)^2-t(t-4m_\pi^2)\right)\left(3B_+^2+2 B_-^2\right)
\label{modm2qpi}
\ee

 The corresponding transition frequency is given by
\be
\omega_{q\pi}(E_a)=\frac{1}{2E_a}\int d\zbf \pi_b f(E_b) W_{ab}^{(q-\pi)}.
\label{omegaqpi}
\ee
 where,
\be
W_{ab}^{(q-\pi)}=\frac{1}{8\pi}\times\frac{1}{2\sqrt{s}p_0}\int dt |\bar M_{q-\pi}|^2(1-f_q)(1+f_\pi)
\label{wabqpi}
\ee
In the above $p_0^2= (s+m_q^2-m_\pi^2)^2/(4s)-m_q^2$. The scattering will contribute to both the
quark relaxation time as well as to the pion relaxation time using Eq. (\ref{omegaqpi}) with appropriate
modification for the initial state.

Let us note that there are poles in the u channel in the quark pion 
scattering term beyond the critical temperature
when the pion mass become larger than the quark mass. However, this is taken care of once we include 
the imaginary part of the quark self-energy in the propagators for the
quarks in the calculation of the amplitude in Eqs. (\ref{bplus})-(\ref{bminus}).
The quark self-energy due to scattering with  mesons can be written as ~\cite{lang}
\be
\Sigma(p_0,\zbf p)=m\Sigma_0+ \zbf\g\cdot\zbf p\Sigma_3-\gamma_0p_0\Sigma_4.
\ee
so that the quark propagators get modified as
\be
S(p_0,\zbf p)=\frac{1}{\diracslash p-m-\Sigma}=\frac{m(1+\Sigma_0)+\gamma_0p_0(1+\Sigma_4)-\zbf \gamma\cdot\zbf p(1+\Sigma_3)}{
p_0^2(1+\Sigma_4)^2-\zbf p^2(1+\Sigma_3)^2-m^2(1+\Sigma_0)^2}.
\ee

The imaginary part of the dimensionless functions $\Sigma_j$, ($j=0,3,4$),i is given as
\be
Im \Sigma_j(p_0,\zbf p)=\frac {g^2}{32 \pi p}d_j\int_{E_{min}}^{E_{max}}
dE_f C_j[f_b(E_b)+f_-(E_f)+f_+(E_f)].
\label{imsig}
\ee
In the above,
$E_b=E_f+p_0$, $p_0=\sqrt{\zbf p^2+m^2}$ and 
$f_\pm$ are the distribution functions for the quarks/antiquarks, $f_b$ is the meson
distribution functions,and,  $C_j$s are  the weight factors given as
\be
C_0=1,\\
C_3=\frac{m_M^2-2m^2-2E_fp_0}{2\zbf p^2},\\
C_4=-\frac{E_f}{p_0}.
\ee

The integration limits are given by
\be
E_{max,min}=\frac{1}{2m^2}\left[(m_M^2-2m^2)p_0\pm |\zbf p|m_M\sqrt{m_M^2-4m^2}\right]
\ee
Further, the degeneracy factors $d_{3,4}$ are 3 for pions and 1 for sigma while $d_0$ is -3 for pions and 1 for 
the sigma meson. To calculate the total relaxation time for a quark of species 'a', we compute the total interaction frequency as 
$\omega_q^{total}(E_a)=\omega(E_a)+\omega_{q\pi}(E_a)$. One can define an average relaxation time for the quarks similar to 
Eq. (\ref{taubarpi}) as 
$\bar{\tau}_q^{total}=\frac{1}{\bar{\omega}_q^{total}}$. 
\be
\bar\omega_q^{total}= \frac{1}{n_q}\int \frac{d\zbf p}{(2\pi)^3}f_q(E)\omega_q^{total}(E)
\label{taubqtot}
\ee

 \begin{figure}
 \begin{center}
 \includegraphics[width=9cm,height=9cm]{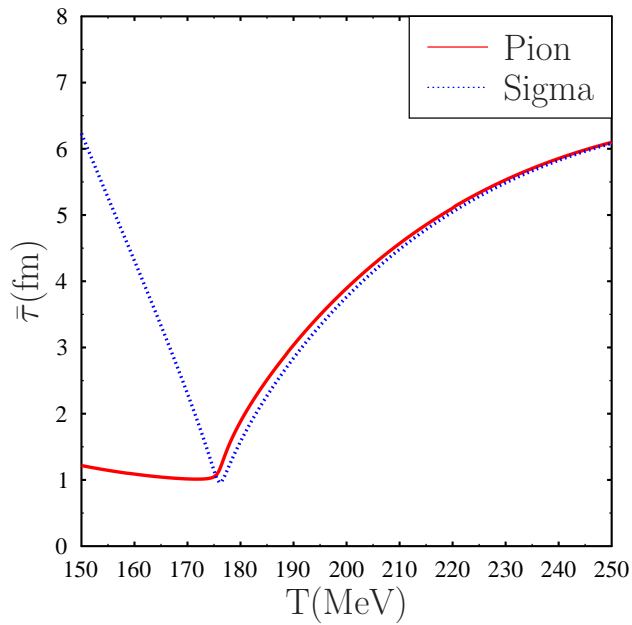}
 \end{center}
\caption {Average relaxation time for pions (solid line) and sigma meson (dotted line). Only meson-meson scatterings are considered here.}
\label{fig5}
\end{figure}

\begin{figure}
 \begin{center}
 \begin{tabular}{c c}
 \includegraphics[width=9cm,height=9cm]{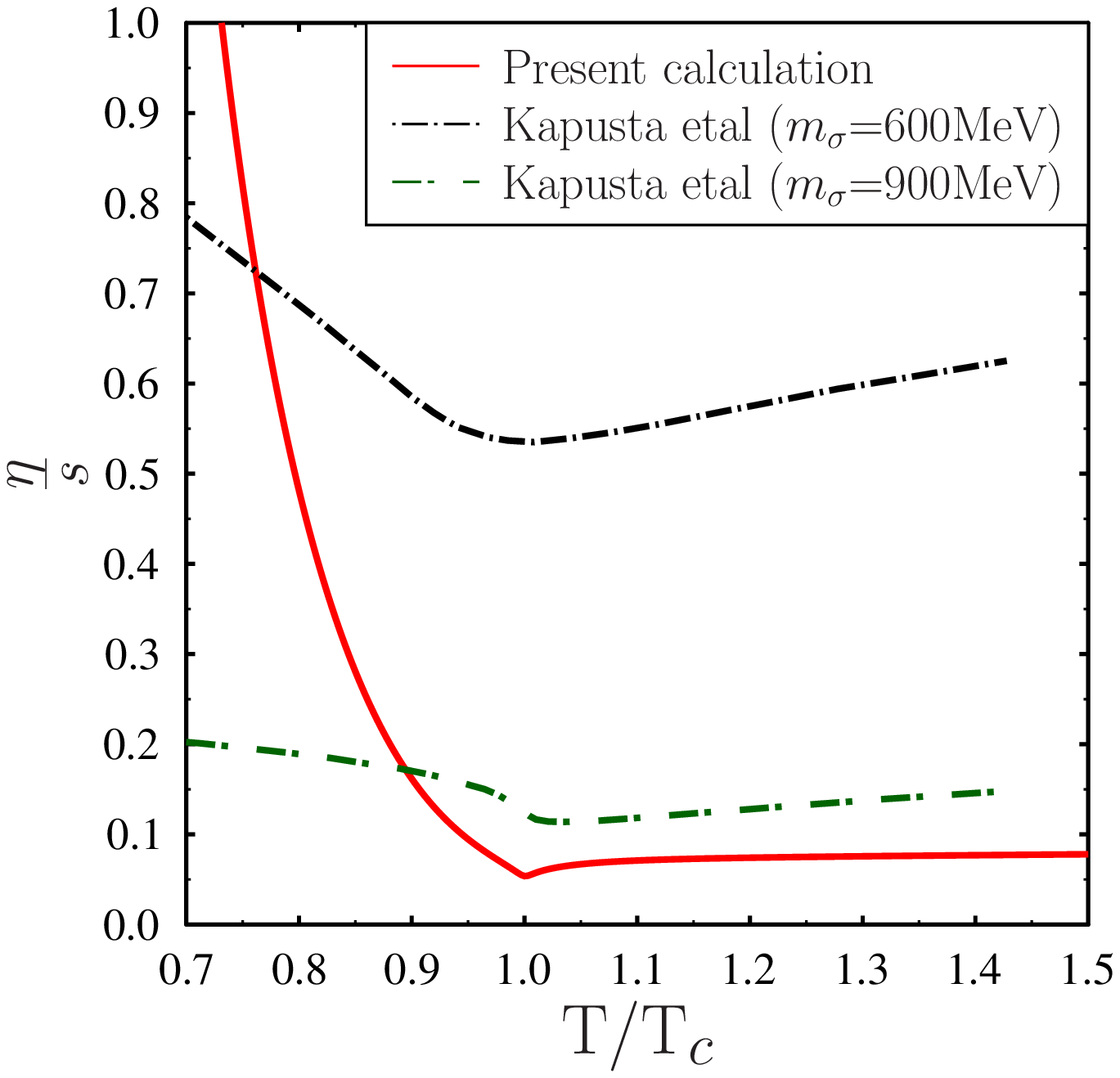}&
 \includegraphics[width=9cm,height=9cm]{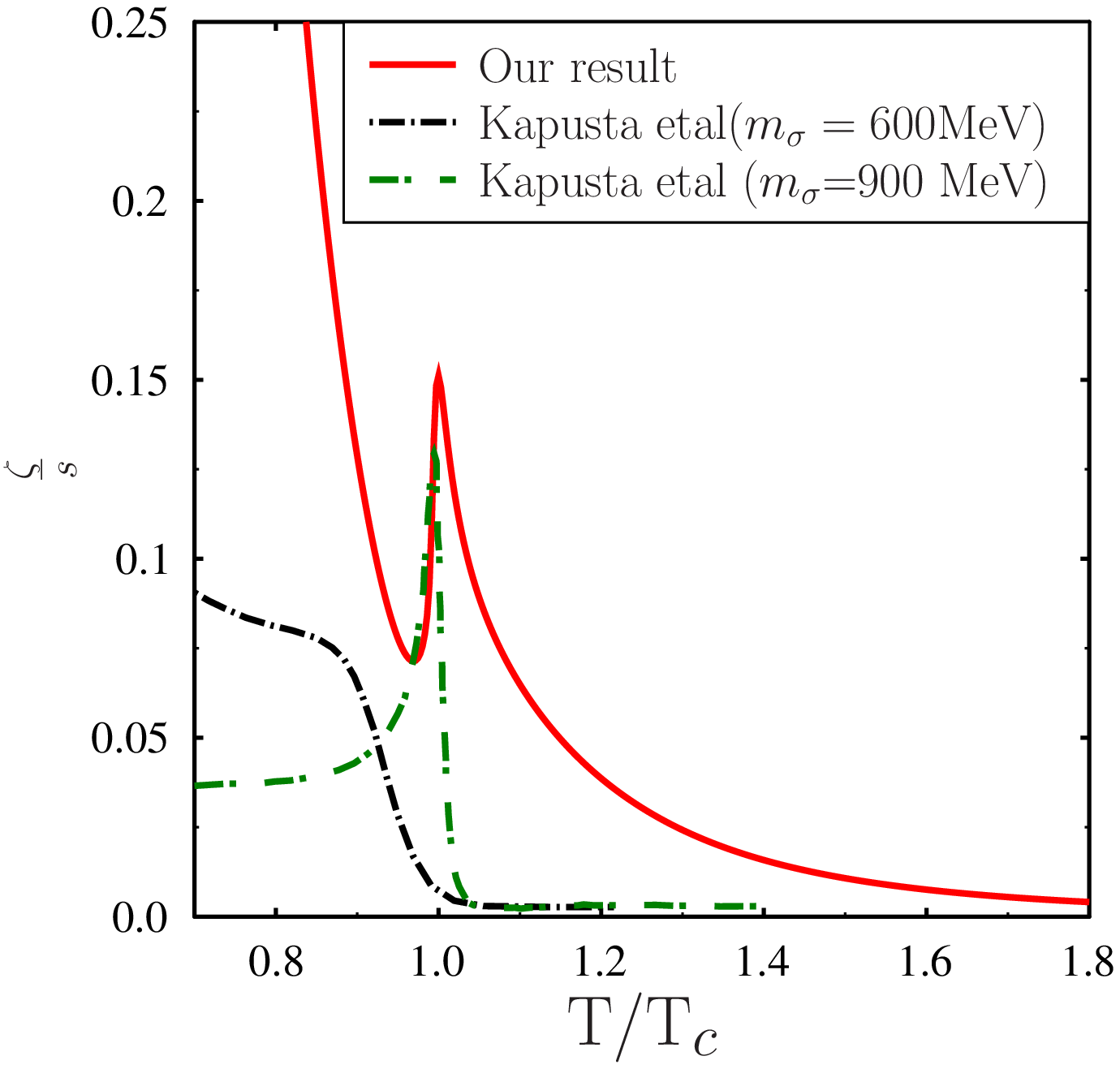}\\
 (a) & (b)
 \end{tabular}
 \end{center}
 \caption{Computations show mesonic contribution calculated using only meson-meson interactions. (a) : Shear viscosity to entropy ratio for $\mu=0$. Present results are shown by solid lines.
The two dot dashed curves correspond to
results of linear sigma model of Ref.~\cite{purnendu} corresponding two different masses for
sigma mesons. (b): Bulk viscosity to entropy ratio for $\mu=0$. Results for current calculations are shown by
solid line. The other results correspond to
 Kapusta et.al. (short dashed) of linear sigma model with ($m_\sigma$=600 MeV),
Kapusta et.al. (dash dot curve) for linear sigma model with $m_\sigma$=900 MeV ~\cite{purnendu}}.

 \label{fig6}
 \end{figure}

\section {Results}
 \subsection{Meson scatterings}
 Let us first discuss the results arising from  meson scattering alone.
Using Eqs. (\ref{omegaea}), with constant $|M|^2$ as discussed, we have plotted the average relaxation times for the $\sigma$-meson
and $\bfm \pi$ mesons in Fig.~\ref{fig5}. The relaxation times are minimum at the transition temperature. Because of larger 
mass of $\sigma$-mesons below the transition temperature, $\bar{\tau}_\sigma$ is much larger  as compared to $\bar{\tau}_{\bfm \pi}$. 
They become almost degenerate after the chiral transition as
may be expected from the behavior of their masses beyond the transition temperature. We may comment here that the particle with larger
relaxation time dominates the viscosities as it can transport energy and momentum to larger distances before interacting. 
 In Fig.~\ref{fig6} we have shown the behavior of the specific  viscosities (normalized to entropy density)
as a function of temperature. In Fig.~\ref{fig6}(a), we have plotted the temperature dependence of the ratio 
$\eta/s$ for $\mu=0$.
 The behavior of this ratio is essentially determined by
 the behavior of the
relaxation time. Similar to Fig.~\ref{fig5}, $\eta/s$ shows a minimum at the crossover temperature and the value at the minimum is
about $\eta/s\sim 0.053$ which is slightly lower than the 
KSS bound of $1/4\pi$. 
 We note that we have considered 
here only the contributions from meson scatterings. As we shall see later,
inclusion of quark degrees of freedom increases the ratio.  
We have also compared with linear sigma model calculations ~\cite{purnendu} in which the quark as well as Polyakov loop
contributions are not taken into account. 
The general behavior of the present calculations is similar to earlier calculations in the sense of 
having a minimum at the chiral crossover temperature. However, the magnitude of the ratio at the critical temperature is
smaller compared to ~\cite{purnendu}.  This is probably due
 to the fact that, the entropy density in the present calculations 
has contributions including those of gluon included through 
the Polyakov loop potential.
The large entropy density, we believe, decreases the magnitude of the ratio. 

In Fig.~\ref{fig6}(b) the ratio of bulk viscosity to entropy  is plotted which shows a maximum at the transition temperature.
We have also plotted in the same figure the results without quarks and Polyakov loop potential.
 The present results show a distinct peak structure in the 
$\zeta/s$ ratio at the crossover temperature. 
 Let us note that
such a peak is expected as an effect of large conformality violation
at the transition temperature as indicated in lattice simulations ~\cite{latticemeyer,karschkharzeev}. 
In Ref.~\cite{purnendu}, a peak structure is seen for a heavier sigma meson ($m_\sigma=900$MeV) which was interpreted as
an effect of stronger self-coupling $\lambda$ for higher $M_\sigma$. However, in the present case, this arises
with quark and polyakov loop degrees of freedom even with a lighter $M_\sigma=600$ MeV.  
 The other characteristic feature  of the present calculation is that, beyond
the critical temperature the ratio $\zeta/s$ falls at a slower rate as compared to  results of previous calculations.
 This has to do with the fact that 
velocity of sound approaches  the ideal gas limit slowly as the effect of Polyakov loops on the quark distribution function
remains significant beyond the critical temperature. In fact, at the transition temperature the value of the Polyakov loop remains about half its value of the ideal limit.
Apart from this, the masses of mesons also get affected by the quark distribution 
functions significantly  beyond the critical temperature. These non ideal effects 
lead to a slower decrease of the ratio beyond the critical temperature.

\begin{figure}
 \begin{center}
  \includegraphics[width=9cm,height=9cm]{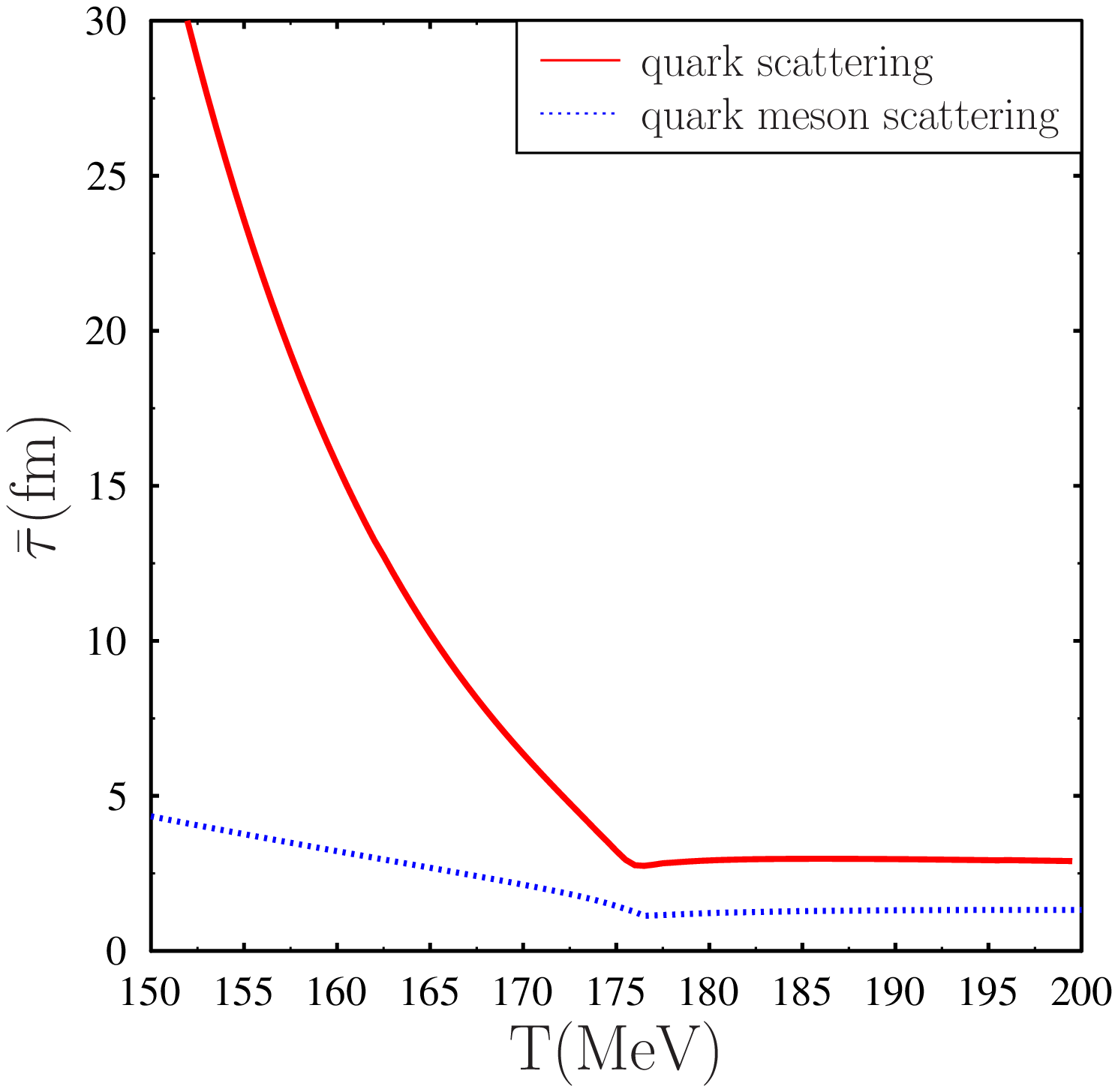}\\
 \end{center}
 \caption {Average relaxation time for quarks arising from quark scattering. The solid curve
corresponds to quark quark/antiquark scatterings with meson exchange. 
The dashed curve
corresponds to including the effect of quark meson scatterings.
Both the curves correspond to $\mu=0$ case.}
 \label{fig7}
 \end{figure}
 
 \subsection{Quark scatterings}
     
 Next, we discuss quark scattering. In Fig.~\ref{fig7} we show the behavior
of average relaxation time for quark scattering. The quark scattering 
through exchange of mesons is shown by the solid line in the figure. Let us recall that the average 
relaxation time is inversely proportional to the transition rate which is related to the cross section.
The dominant contribution here comes from the quark-antiquark scattering from the $s$ channels through
propagation of the resonance states, the pions and the sigma mesons. The masses of the sigma meson
decrease with temperature, becoming a minimum at the transition temperature,
leading to an enhancement of the cross section. Beyond this, the cross section decreases due to the increase in the masses of the mesons. This,
in turn, leads to a minimum in the relaxation time. 

The average relaxation time  for quarks including the  quark meson scattering  along with the quark scattering
 is shown as the dashed curve in Fig.~\ref{fig7}. This curve lies below the quark quark scattering curve as
there is additional contribution to the transition rate from the quark meson scattering. Below the critical temperature, the quark meson scattering dominates over the quark quark scattering due to the smaller
mass of the pions as compared to the massive constituent quarks. Beyond the critical temperature,
one would have expected the quark meson scattering contribution to be negligible because of the
suppression due to the large meson masses. However, as was noted earlier, beyond the critical temperature,
there are poles in the scattering amplitude in the $u$-channel for quark-pion scattering as the
pion mass becomes larger than the quark masses. This is, however, regulated by the finite width of the
quarks as calculated in Eq. (62). Nonetheless, the contribution of the quark pion 
scattering to the total quark interaction frequency $\omega_{q\pi}$(E) is non-negligible beyond
 the critical temperature.

\begin{figure}
 \begin{center}
 \begin{tabular}{c c}
 \includegraphics[width=9cm,height=9cm]{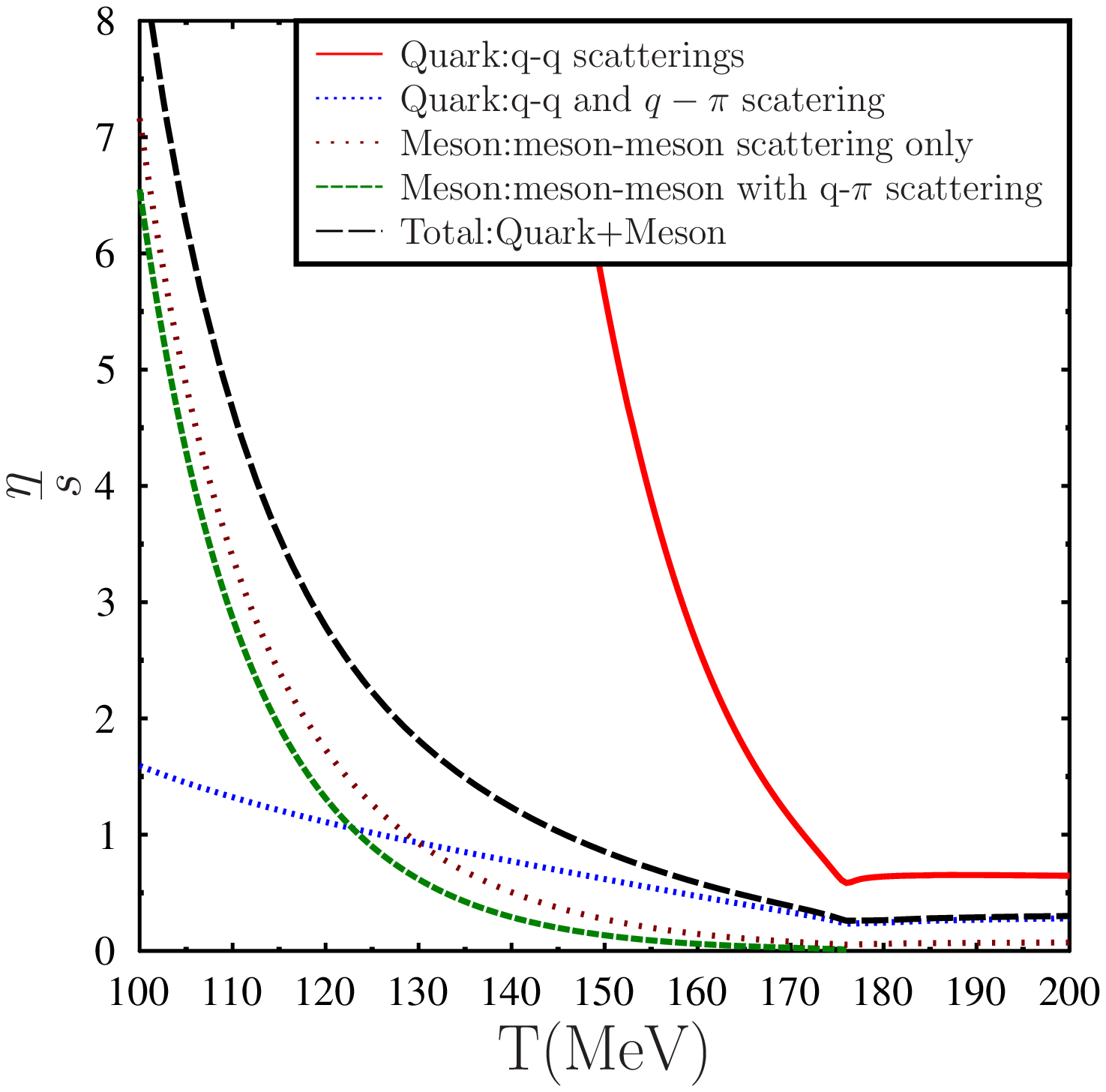}&
 \includegraphics[width=9cm,height=9cm]{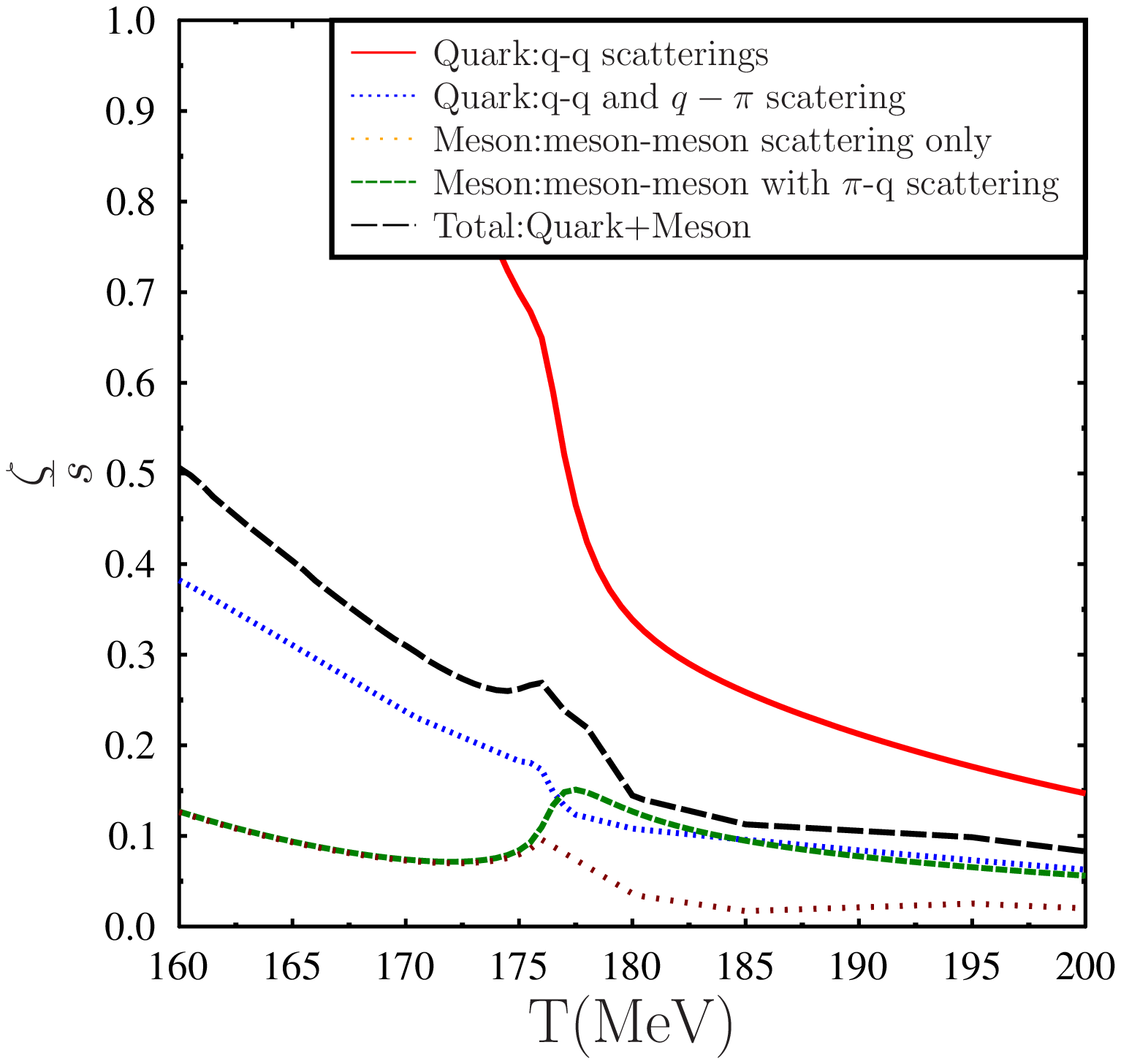}\\
 (a) & (b)
 \end{tabular}
 \end{center}

 \caption { Different contributions for specific viscosity coefficients.
$\eta/s$ is shown in the left while $\zeta/s$ is shown on the right.
In both the figures, contributions from the quarks with relaxation time computed using only quark-quark scattering(red solid line) 
and also including quark-meson scattering(blue dotted line) are shown as a function of temperature.
 The contribution of the mesons due to meson-meson scattering
 (green dashed curve) and including meson-quark scattering (maroon short dashed curve)
is also shown. The total contribution from the quarks and mesons are
is shown by the black long dashed curve.
All the curves correspond to $\mu=0$ case.}
 \label{fig8}
 \end{figure}

 We next discuss the contribution of different scatterings to the 
specific shear viscosity $\eta/s$. The same is shown in Fig.~\ref{fig8}(a) for
vanishing chemical potential. The contribution from the mesons to the shear viscosity is arising from
 the meson scatterings only is shown by the green dashed curve while the effect of including the meson-quark scattering is shown by the maroon dotted curve. Similarly the quark contribution to this ratio $\eta/s$ arising from quark quark scattering only is shown by the red solid line while the total
contributions including the quark-pion scattering is shown by the blue dotted line. This also demonstrates the importance of the scattering of quarks and mesons to the total viscosity coefficient. The total contributions from both the quarks and mesons is shown as the black dashed curve in Fig.~\ref{fig8}.

In a similar manner,  various contributions to the specific bulk 
viscosity ($\zeta/s$) coefficient are shown in Fig.~\ref{fig8}(b). 
As may be 
observed, while no peak structure is seen for this coefficient from 
the contributions arising from quarks scatterings only, such a structure is seen only when one includes the quark meson scattering. 
The total effect is shown as black dashed curve in Fig.~\ref{fig8}(b).
      
\begin{figure}
 \begin{center}
 \begin{tabular}{c c}
 \includegraphics[width=9cm,height=9cm]{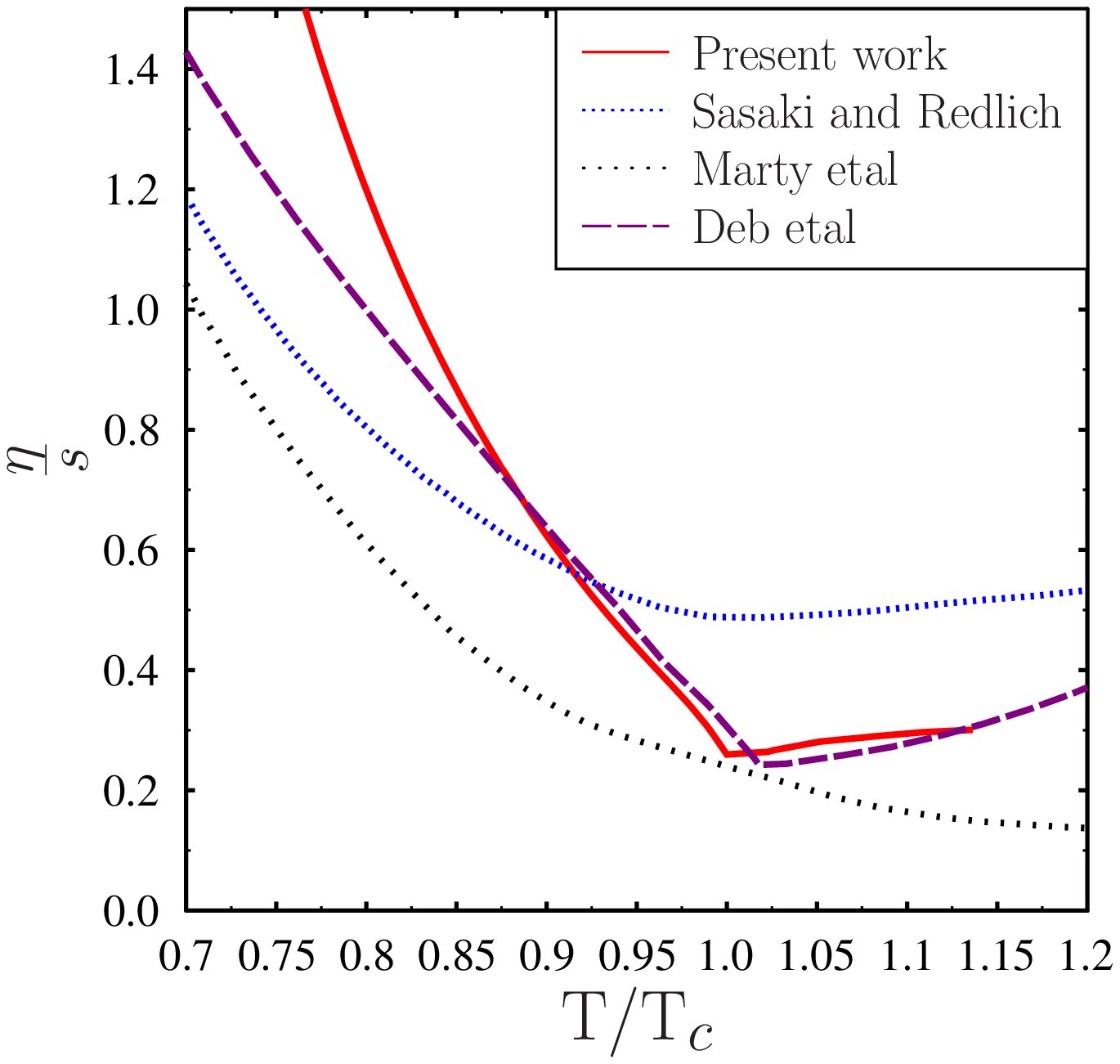}&
 \includegraphics[width=9cm,height=9cm]{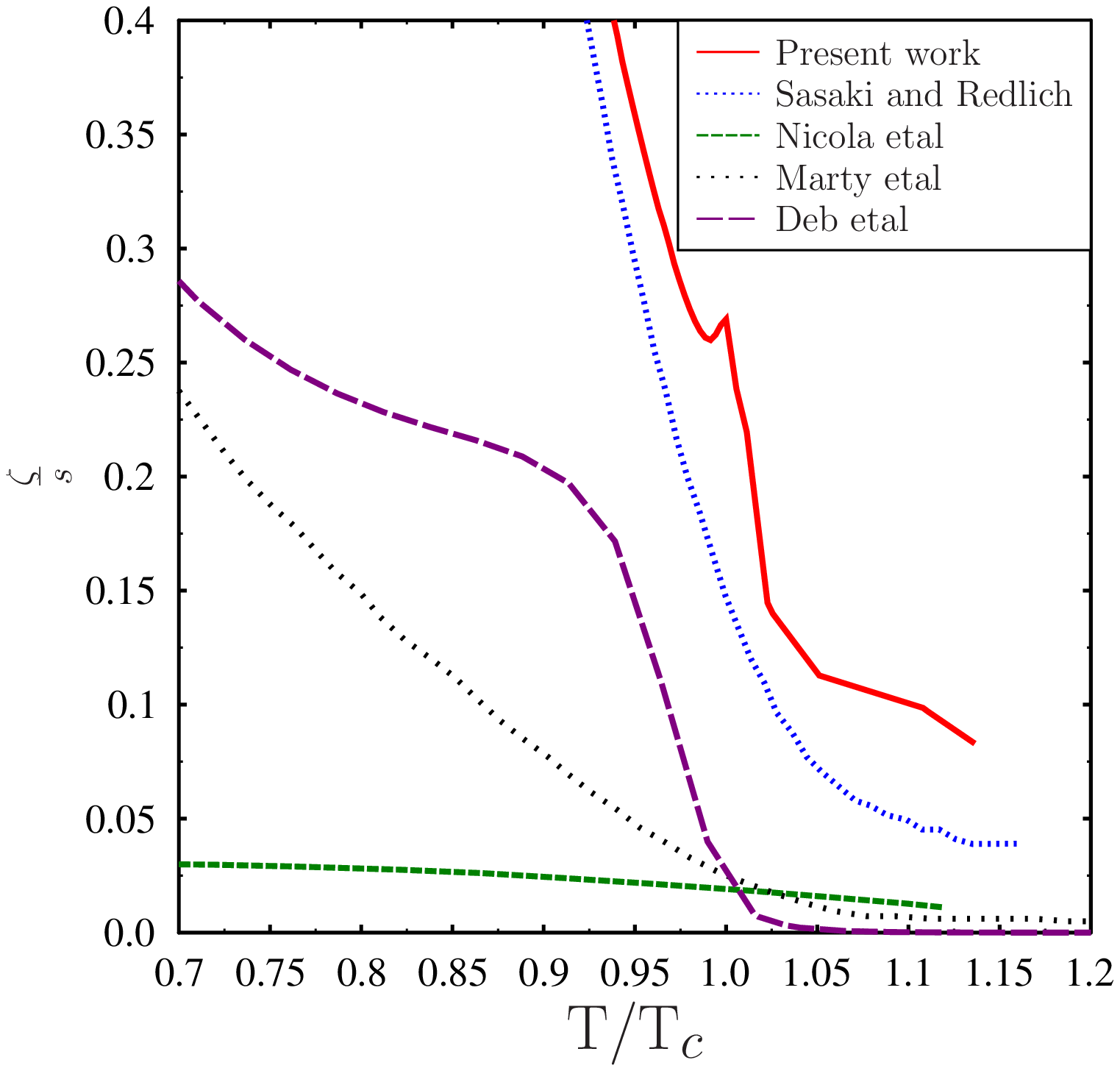}\\
 (a) & (b)
 \end{tabular}
 \end{center}
 \caption{ (a) : Shear viscosity to entropy ratio for $\mu=0$. Present results are shown by solid lines.
The dotted line correspond to results of NJL model of Ref.~\cite{sasakinjl}
, the short dashed curve correspond to results of Marty et.al. Ref.~\cite{marty}
and the long dashed curves correspond to results of Deb. et.al. Ref.~\cite{deb}.
(b): The results of Bulk viscosity to entropy ratio compared with other
 results in NJL models. The notation is similar to of (a).}
 \label{fig9}
 \end{figure}

In Fig.~\ref{fig9}, we compare the present results with earlier works on the NJL model.
As may be noted, in general, the behavior is similar regarding the
shear viscosity-to-entropy ratio. Both NJL as well as the present calculations
of the PQM model show the similiar behavior of having a minimum at the transition
temperature as in Refs. ~\cite{deb,sasakinjl}. The results of Ref. ~\cite{marty}, 
on the other hand, show a monotonic decrease with temperature. The bulk viscosity-to-entropy ratio, 
here however shows a much faster rise as the temperature
is lowered below the critical temperature. In fact, both the specific viscosities rise much faster compared to NJL models below the critical temperature in the PQM model considered here. The reason could be due to the fact that the entropy
density for PQM model is smaller compared to NJL models. The Polyakov loop
decreases as temperature is lowered which leads to a suppression of quark distribution functions leading to decrease of entropy density at a faster rate as compared to NJL model. Moreover, within the present approximation pions do not
contribute to the thermodynamics here. Further, for temerature  larger than the critical temperature, the bulk viscosity vanishes slowly with increase in
temperature as compared to NJL model. This is due to the fact that the Polyakov loop variable takes its asymptotic values only at very high temperatures. 

\begin{figure}
 \begin{center}
  \includegraphics[width=9cm,height=9cm]{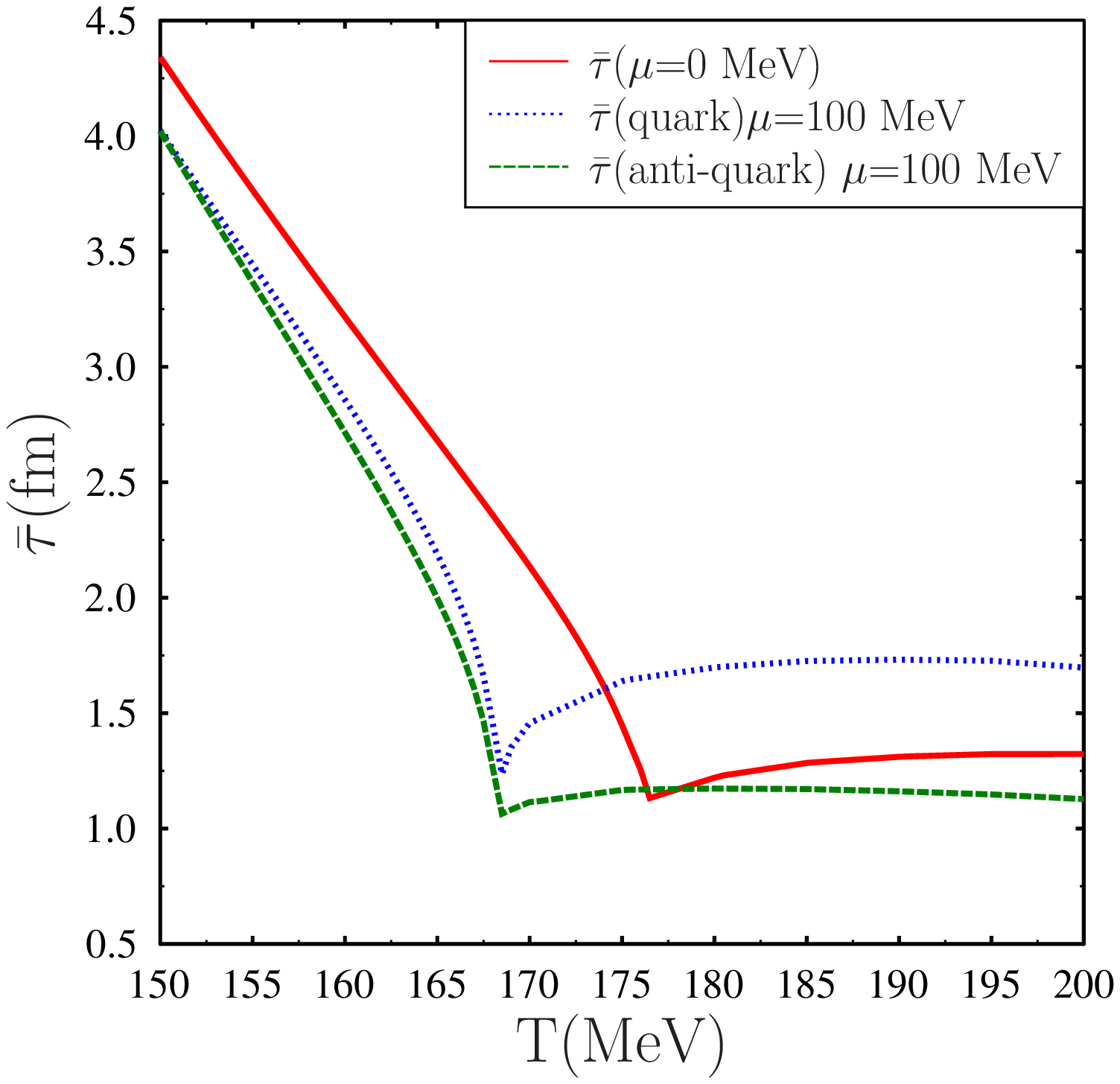}\\
 \end{center}
 \caption {Average relaxation time of quarks and antiquarks for $\mu=100$ MeV.
The solid line correspond to the case of  $\mu=0$ MeV. }
 \label{fig10}
 \end{figure}

\begin{figure}
 \begin{center}
 \begin{tabular}{c c}
 \includegraphics[width=9cm,height=9cm]{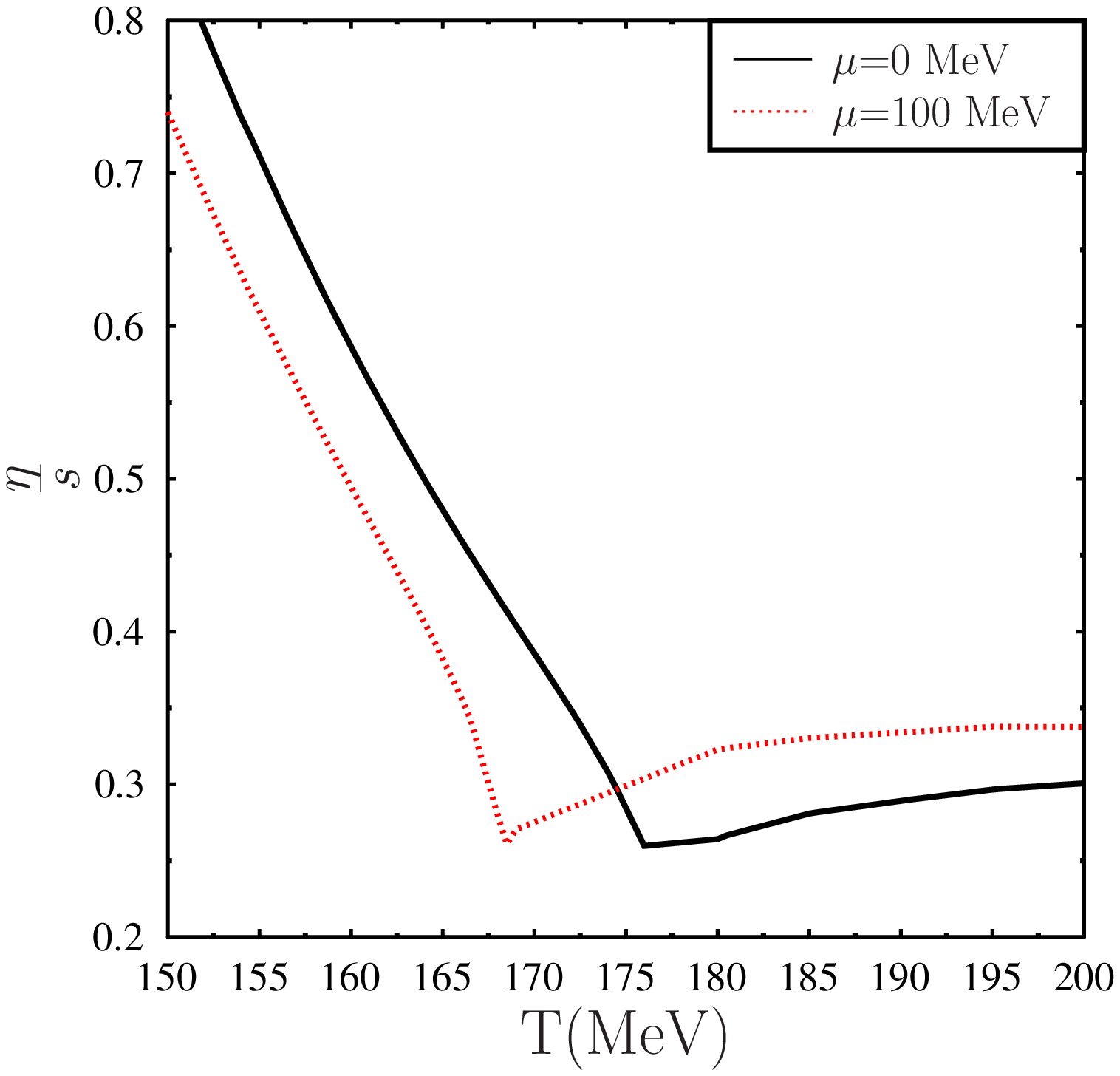}&
 \includegraphics[width=9cm,height=9cm]{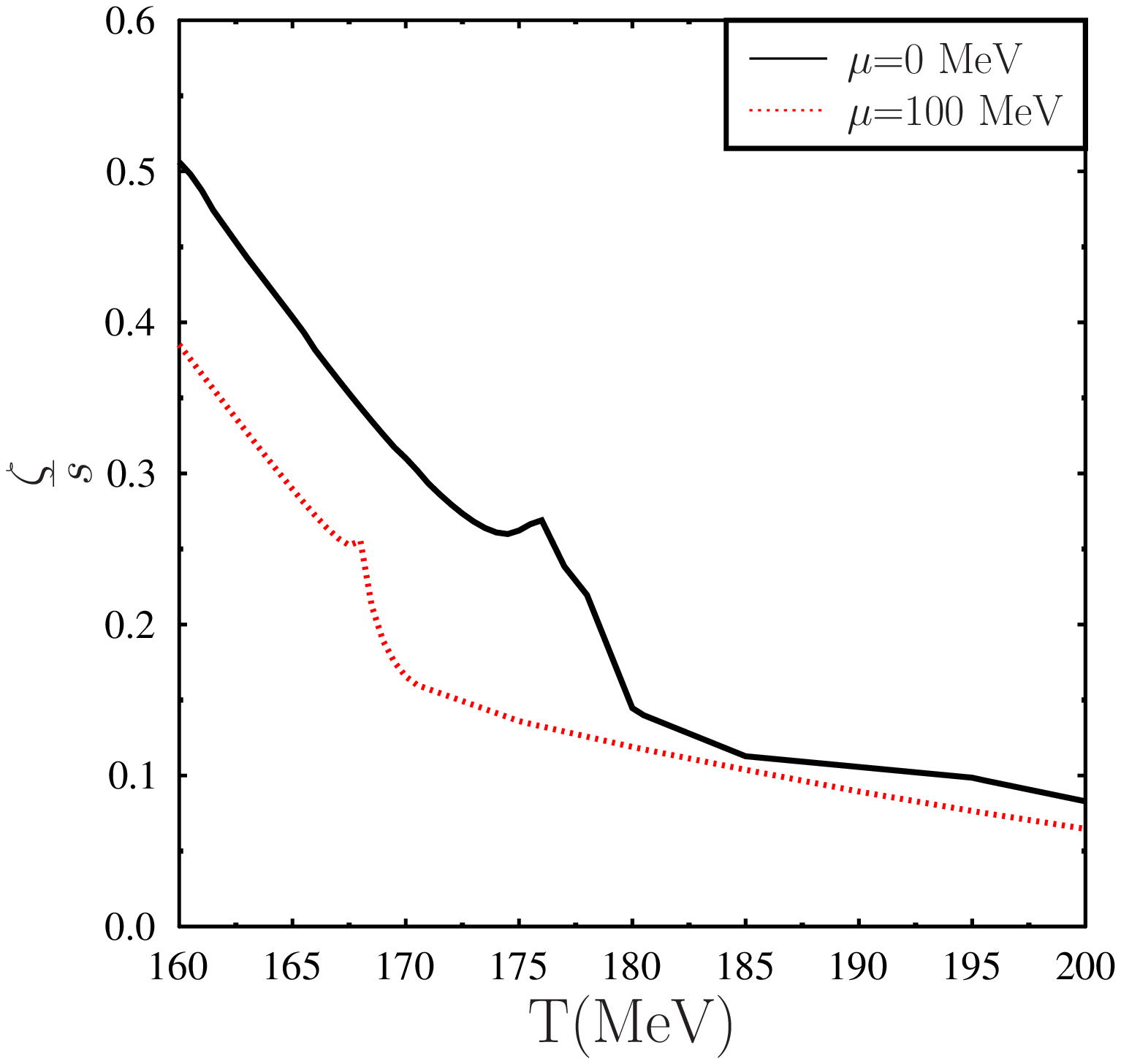}\\
 (a) & (b)
 \end{tabular}
 \end{center}
 \caption{Viscosities for $\mu=100$ MeV. The left figure shows $\eta/s$ as a function of temperature
for $\mu=0$ MeV (solid line) and $\mu=100$ MeV (dotted line). 
The right figure shows the
ratio $\zeta/s$ as a function of temperature.}
 \label{fig11}
 \end{figure}

\begin{figure}
\includegraphics [width=8cm,height=8cm]{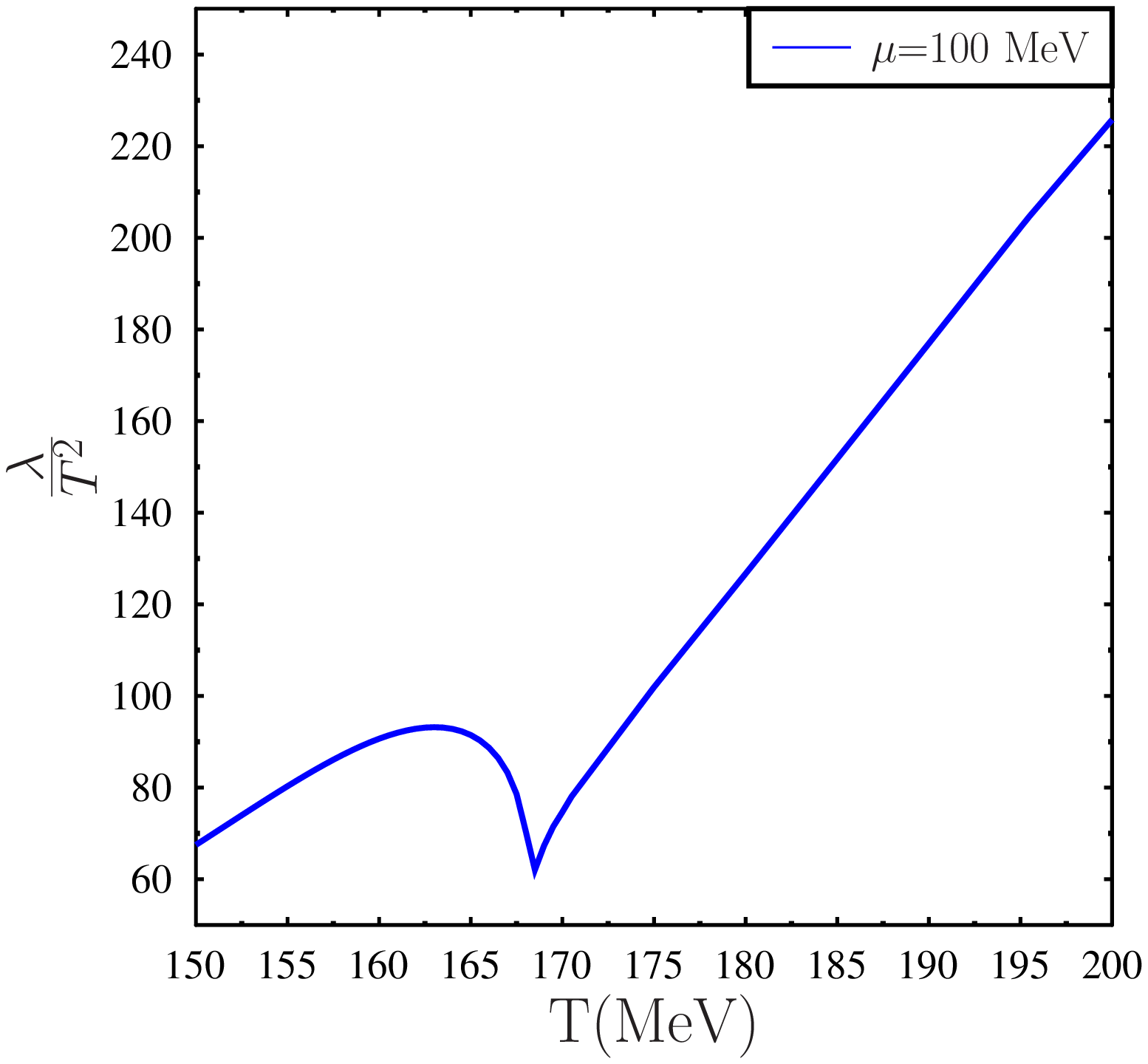}
\caption{Thermal conductivity in units of $T^2$ as a function of temperature
for $\mu=100$MeV.   }
\label{fig12}
\end{figure}

Next, we discuss about effect of finite chemical potential on the 
transport coefficients. To begin with let us note that the average relaxation time $\bar{\tau}_a$ as in Eq. (\ref{taubqtot}) 
depends both on the transition rate and the density of the particles in the initial state. To this end,
 let us discuss the case of T$>$T$_c$. Here, the quark densities are larger than those of antiquarks. Further, the 
dominant contribution in this range of temperatures arises from $u \bar{d} \rightarrow u \bar{d}$ scatterings.
 As there are fewer antiquarks to scatter off, the average transition frequency of quark-antiquark scattering decreases. 
This leads to $\bar{\tau}_q(\mu)>\bar{\tau}_q(\mu=0)$. On the other hand, for the antiquarks, there are more quarks 
to scatter off than compared to the case of $\mu=0$. Hence, this leads to $\bar{\tau}_{\bar{q}}(\mu)<\bar{\tau}_{\bar{q}}(\mu=0)$. 
This expected behavior is seen in Fig.~\ref{fig10}. Next, let us consider the case T$<$T$_c$. In this case, the antiquark density 
is heavily suppressed due to constituent quark mass and the chemical potential and dominant contribution for quark relaxation time, therefore
 arises from quark-quark scatterings. This leads to $\bar{\tau}_{q}(\mu)<\bar{\tau}_{q}(\mu=0)$. On the other hand, for the antiquarks, 
though their number density is smaller, their interaction frequency is enhanced both by the larger amplitude for
 M$_{u \bar{d} \rightarrow u \bar{d}}$ scattering and the larger number of quarks as compared to case at $\mu$=0. 
This leads to $\bar{\tau}_{\bar{q}}(\mu)<\bar{\tau}_{q}(\mu=0)<\bar{\tau}_{\bar{q}}(\mu=0)$. This general behavior is 
reflected in the average relaxation time dependence on T in Fig.~\ref{fig10} below  the critical temperature.

In Fig.~\ref{fig11}, we have shown the results for the viscosities at $\mu=100$ MeV.
Fig.~\ref{fig11} (a) shows the variation of the specific shear viscosity ($\eta/s$) as a function of temperature for zero and 
finite chemical potential. The behavior of shear viscosity
 essentially follows that of the  behavior of the relaxation time. 
$\eta/s$ has a 
minimum at the critical temperature with $\eta/s|_{min}\sim 0.23$ ($\mu=0$)
 due to suppression of the scattering cross section at higher temperature.
 At finite $\mu$, the ratio is little higher as compared to the 
 value at vanishing $\mu$. This is due to two reasons. Firstly, 
the relaxation time at nonzero chemical potential is larger and,
 moreover, the quark density also becomes larger at finite chemical potential.  
At temperatures below the critical temperature and near the critical
 temperature,$\eta/s(\mu)<\eta/s(\mu=0)$ as the relaxation time is
 lower. However, at lower temperatures, the meson scattering 
becomes significant and $\eta/s$ 
for finite chemical potential becomes similar to that at vanishing 
chemical potential as is observed in the figure.

In Fig.~\ref{fig11}(b), we have plotted the bulk viscosity-to-entropy ratio 
for $\mu=0$ MeV and $\mu=100$ MeV. It turns out that at finite $\mu$ 
the specific bulk viscosity
is smaller than the value at $\mu=0$ MeV. The reason for it is the fact that 
the dominating contribution to the finite $\mu$ arise from the term $M^2-TM\frac{dM}{dT}-\mu M\frac{dM}{d\mu}$ in the expression for $\zeta/s$ in Eq. (\ref{zeta}). This is due to the sharp variations of the order parameters at finite chemical potential as may be observed in Fig.~\ref{fig3}. As this term contributes
 negatively to the expression for $\zeta$, the specific bulk viscosity at finite $\mu$ is lower than that at $\mu=0$ MeV.

In Fig.~\ref{fig12} , we have shown the results for thermal conductivity. We have plotted here the dimensionless quantity $\lambda/T^2$ as 
a function of temperature. We have plotted the results for $\mu=100 $MeV. As is well known,
thermal conduction which involves the relative flow of energy and baryon number vanishes at zero baryon density. In fact,
$\lambda$ diverges as $1/n^2$ as may be expected from the expression given in Eq. (\ref{lamb}). However, in the dissipative current,
 the conductivity occurs as $\lambda n^2$ ~\cite{gavin,hosoya} and the heat conduction vanishes for $\mu=0$ ~\cite{daniel}.  On the other hand, in some cases, such as when pion number is conserved, heat conduction can be sustained by pions. In presece of a 
pionic chemical potential corresponding to a conserved pion number, 
  thermal conductivity can be  nonzero at vanishing baryonic chemical potential. This has been the 
basis for estimation of thermal conductivity at zero baryon density but  finite pion density ~\cite{nicola,sabyath,souravsuk}. 
However, in the present case, we consider the case of vanishing  pion chemical potential and 
 show only the contribution of quarks to thermal conductivity.

As expected from the behavior of the relaxation time, the specific thermal conductivity has a minimum at 
the critical temperature similar to Ref.~\cite{deb} for the  NJL
model. The sharp rise of $\lambda/{\text T}^2$ can be understood by
performing  a dimensional argument to show that at very high temperature when
chiral symmetry is is restored the integral increases
as $\text T^3$ while the prefactor $w/(nT)$
grows as $\text T^2$ for small chemical potentials. Apart from this
kinematic consideration, the integrand further is multiplied
 by $\tau(E)$ which itself is an increasing function of temperature
beyond $T_c$. This leads to the sharp rise of the
ratio $\lambda/T^2$ beyond the critical temperature. Below, the critical
temperature, however, the ratio decreases which is in contrast to NJL results
of Ref.~\cite{deb}. The reason is twofold. First, the magnitude of the relaxation time decreases when quark meson scattering 
is included as compared to quark-quark scattering as shown in Fig.~\ref{fig7}. Apart from this, 
in the integrand,the distribution functions are suppressed by Polyakov loops as compared to NJL model.
 As the antiquark densities are suppressed compared to quark densities at finite chemical potential,
the high-temperature behavior is decided by the quark-quark scattering.

\section*{Summary}

Transport coefficients of hot and dense matter are important inputs
for the hydrodynamic evolution of the plasma that is produced following a heavy ion collision. In the present study, we have investigated these cofficients
taking into account the the nonperturbative effects related to chiral
 symmetry breaking as well as confinement properties of strong interaction physics within
an effective model, the Polyakov loop extended quark meson coupling model.
These coefficients are estimated using the relaxation time approximation
for the solutions of the Boltzman kinetic equation.

We first calculated the medium-dependent masses of the mesons and quarks within
a mean field approximation. The contribution of the mesons to the 
transport coefficients has been calculated through estimating 
the relaxation time for the mesons arising both from meson-meson scattering 
and meson-quark scattering. The 
contribution to the transport coefficients arises mostly from the meson 
scatterings at temperatures below the critical temperature,
 while above the critical temperature, the contributions arising 
from the quark scatterings become dominant. In particular, 
quark-meson scattering contributes significantly to the
relaxation time for the quarks both below and above the critical temperature. 
The quark-pion scattering above the critical temperature gives significant 
contribution due to the pole structure of the corresponding 
scattering amplitude. 

One important approximation in the present analysis is that the kinetic terms for the mesons are not modified at finite temperature
 and meson dispersion relation remains similar to those at the zero-temperature relativistic dispersion relation. The only temperature
 effect that remains in the meson dispersion lies in the temperature-dependent meson masses obtained through 
the curvature of the effective potential  ~\cite{rischkeqm}. A more realistic approach would be to use effective
 field theory to have different dispersion relations for the mesons ~\cite{pion} depending upon their velocities
 and calculate the scattering processes 
to estimate the viscosities. However, such an approach is beyond the scope of present work in which we have restricted ourselves
 to thermal and density effects included in the masses and widths for the mesons.

In general, the effect of Polyakov loops lies in suppressing the quark 
contribution below the critical temperature. This leads to, in particular,
 the suppression of thermal conductivity at lower temperature arising 
from quark scattering.
The effect of Polyakov loop also is significant near and above the critical
temperature. Indeed, both the quark masses as well as Polyakov loop
order parameter remain significantly different from their
asymptotic values  near the critical temperature.
It will be interesting to examine the consequences of such nonperturbative 
features on the
transport coefficients of heavy quarks as well as on the 
collective modes of QGP
above and near the critical temperature. Some of these works are in progress and will be reported elsewhere.

\acknowledgements
The authors would like to acknowledge many discussions with Guru Prasad Kadam 
and Pracheta Singha. SG is financially supported by University Grants 
Commission Dr. D.S. Kothari Post Doctoral Fellowship (India),
 under grant no. F4-2/2006 (BSR)/PH/15-16/0060.

\def\berera{A. Bstero-Gil, A. Berera and R. Ramos, JCAP1107, 030 (2011).}
\def\heinzrev{U. Heinz and R. Snellings, Annu. Rev. Nucl. Part. Sci. 63, 123-151, 2013}
\def\pethick{H. Heiselberg and C. Pethick,{\PRD{48}{2916}{1993}}.}
\def\exptstar{J. Adams,{\em et al}(STAR),{\NPA{757}{102}{2005}}.}

\def\karschkharzeev{F. Karsch, D. Kharzeev, and K. Tuchin, Phys. Lett. B
663, 217 (2008).}
\def\kharzeevtuchin{D. Kharzeev, and K. Tuchin,JHEP 0808,031, (2008).}
\def\nicolaprl{	
D. Fernandez-Fraile, A. Gomez Nicola, {\PRL{102}{121601}{2009}}.}
\def\joglekar{J.C. Collins, A. Duncan, S.D. Joglekar, Phys. Rev. D 16, 
438 (1977).}
\def\blaschke{J. Jankowski, D. Blaschke, M.Spalinski, Phys.Rev.D 87, 105018^M
(2013). }
\def\gorenstein{M. Gorenstein, M. Hauer, O. Moroz, Phys.Rev.C 77,024911 (2008)}
\def\bugaev{K. Bugaev et al, Eur.Phys.J. A 49, 30 (2013)}
\def\cpsingh{S.K. Tiwari, P.K. Srivastava, C.P. Singh, Phys.Rev. C 85,
014908 (2012) }
\def\chen{J.-W. Chen, Y-H. Li, Y.-F. Liu, and E. Nakano, Phys. Rev. D 76,
114011 (2007)}
\def\chennakano{J.-W. Chen, and E. Nakano, Phys. Lett. B 647, 371 (2007)}
\def\itakura{K. Itakura, O. Morimatsu, and H. Otomo, Phys. Rev. D 77, 014014
(2008)}
\def\cleymans{J. Cleymans, H. Oeschler, K. Redlich, and S. Wheaton, Phys.
Rev. C 73, 034905 (2006)}
\def\worku{J. Cleymans and D. Worku, Mod. Phys. Lett. A26,1197,(2011).}
\def\guptagod{S. Chatterjee, R. M. Godbole and S. Gupta, {\PRC{81}{044907}{2010}}.}
\def\Noronha{Noronha-Hostler J, Noronha J and Greiner C 2012 Phys. Rev. C 86 024913}
\def\tanmoy{A. Bazavov {\it et.al.}, {\PRD{90}{094503}{2014}}}
\def\cavitation{K. Rajagopal and N. Trupuraneni, JHEP1003, 018(2010);
 J. Bhatt, H. Mishra and V. Sreekanth, JHEP 1011, 106,(2010);{\it ibid} Phys. Lett. B704, 486 (2011); {\it ibid} Nucl. Phys. A875, 181(2012).}
\def\borsonyi{S. Borsonyi{\it et.al.}, JHEP1011, 077 (2010).}
\def\borsonyimu{S. Borsonyi{\it et.al.}, JHEP1208, 053 (2012).}
\def\dobadoch{A. Dobado,F.J.Llane-Estrada amd J. Torres Rincon, 
 J. Bhatt, H. Mishra and V. Sreekanth, JHEP 1011, 106,(2010);{\it ibid} Phys. Lett. B704, 486 (2011); {\it ibid} Nucl. Phys. A875, 181(2012).}
\def\borsonyi{S. Borsonyi{\it et.al.}, JHEP1011, 077 (2010).}
\def\borsonyimu{S. Borsonyi{\it et.al.}, JHEP1208, 053 (2012).}
\def\dobadoch{A. Dobado,F.J.Llane-Estrada amd J. Torres Rincon, 
{\PLB{702}{43}{2011}}.}
\def\dobadoshear{A. Dobado,F.J.Llane-Estrada amd J. Torres Rincon, 
{\PRD{79}{055207}{2009}}.}
\def\sasakiqp{C. Sasaki and K.Redlich,{\PRC{79}{055207}{2009}}.}
\def\sasakinjl{C. Sasaki and K.Redlich,{\NPA{832}{62}{2010}}.}
\def\ellislet{I.A. Shushpanov, J. Kapusta and P.J. Ellis,{\PRC{59}{2931}{1999}}
; P.J. Ellis, J.I. Kapusta, H.-B. Tang,{\PLB{443}{63}{1998}}.}
\def\prakashwiranata{Anton Wiranata and Madappa Prakash,
{\PRC{85},{054908}{2012}}.}
\def\purnendu{P. Chakraborty and J.I. Kapusta {\PRC{83}{014906}{2011}}.}
\def\greco{S.Plumari,A. Paglisi,F. Scardina and V. Greco,{\PRC{83}{054902}{2012}a.}}
\def\bes{H. Caines, arXiv:0906.0305 [nucl-ex], 2009.}
\def\greinerprl{J. Noronha-Hostler,J. Noronha and C. Greiner,
{\PRL{103}{172302}{2009}}.}
\def\greinerprc{J. Noronha-Hostler,J. Noronha and C. Greiner
, {\PRC{86}{024913}{2012}}.}
\def\igorgreiner{J. Noronha-Hostler, C. Greiner and I. Shovkovy,
, {\PRL{100}{252301}{2008}}.}
\def\majumdermueller{A. Majumder and B. Mueller, {\PRL{105}{252002}{2010}}.}
\def\leonidov{ A. V. Leonidov and P. V. Ruuskanen, {\EPJC{4}{519}{1998}}.}
\def\cbm{ B. Friman, C.H. Ohne, J. Knoll, S. Leupold, J. Randrup, R. Rapp, P. Senger (Eds.), Lect. Notes Phys., vol. 814,
2011.}
\def\nica {A.N. Sissakian, A.S. Sorin, J. Phys. G 36 (2009) 064069.}
\def\nakano{J.W. Chen,Y.H. Li, Y.F. Liu and E. Nakano,
 {\PRD{76}{114011}{2007}}.}
\def\wang{M.Wang,Y. Jiang, B. Wang, W. Sun and H. Zong, Mod. Phys. lett.
{\bf A76}, 1797,(2011).}
\def\agasian{N.O. Agasian, JETP Lett. 95, 171, (2012), arXiv:1109.5849.}
\def\Hagedorn{R. Hagedorn and J. Rafelski,{\PLB{97}{136}{1980}}.}
\def\kapustaolive{J.I. Kapusta and K. A. Olive, {\NPA{408}{478}{1983}}.}
\def\hrgexp{P. Braunmunzinger, J. Stachel, J.P. Wessels and N. Xu,
{\PLB{365}{1}{1996}}; G.D. Yen and M.I. Gorenstein, {\PRC{59}{2788}{1999}};
F. Becattini, J. Cleymans, A. Keranen, E. suhonen and K. Redlich, 
{\PRC{64}{024901}{2001}}.}
\def\rischkegorenstein{.D.H. Rischke, M.I. Gorenstein, H. Stoecker and
W. Greiner, Z.Phys. C {\bf 51}, 485 (1991).}
\def\hmnjl{Amruta Mishra and Hiranmaya Mishra, {\PRD{74}{054024}{2006}}.}
\def\pdgb{C. Amseler {\it et al}, {\PLB{667}{1}{2008}}.}
\def\shuryak{E.V. Shuryak, Yad. Fiz. {\bf 16},395, (1972).}
\def\leupold{S. Leupold, J. Phys. G{\bf32},2199,(2006)}
\def\peter{A. Andronic, P. Braun-Munzinger , J. Stachel and M. Winn,
{\PLB{718}{80}{2012}}}
\def\blum{M. Blum, B. Kampfer, R. Schluze, D. Seipt and U. Heinz,{\PRC{76}{034901}{2007}}.}
\def\jaminplb{M. Jamin, {\PLB{538}{71}{2002}}.}
\def\ghosh{Sabyasachi Ghosh, {\PRC{90}{025202}{2014}}; International Journal Of Modern Physics {\bf A29}, 145005,2014.}
\def\csernai{L.P. Csernai, J.I. Kapusta and L.D. McLerran,{\PRL{97}{152303}{2006}}.}
\def\hagedorn{R. Hagedorn, Nuovo Cim. Suppl. 3,147 (1965); Nuovo Sim. A56,1027 (1968).}
\def\torieri{G. Torrieri and I. Mishustin,{\PRC{77}{034903}{2008}}.}
\def\fernandez{D. Fernandiz-Fraile and A.G. Nicola,{\PRL{102}{121601}{2009}}.}
\def\caron{S.Caron,{\PRD{79}{125009}{2009}}.}
\def\latticemeyer{H.B. Meyer,{\PRL{100}{162001}{2008}}.}
\def\romatschke{P.Romatscke and D.T. Son,{\PRD{80}{065021}{2009}}.}
\def\moore{G.D. Mooore and O. Sarem, J. High Energy Phys. JHEP0809(2008)015.}
\def\dobado{A.Dobado and J. M. Torres-Rincon {\PRD{86}{074021}{2012}}.}
\def\monai{A. Monnai and T. Hirano, Phys. Rev. C 80, 054906 (2009).}
\def\kodama{G. S. Denicol, T. Kodama, T. Koide, and P. Mota, Phys. Rev. C 80, 064901 (2009).}
\def\heinz{H. Song and U. Heinz, Phys. Rev. C 81, 024905 (2010).}
\def\bozek{P. Bozek, Phys. Rev. C 81, 034909 (2010).}
\def\schaferdus{K. Dusling and T. Schafer,  ̈ Phys. Rev. C 85, 044909 (2012).}
\def\noronhahydro{J. Noronha-Hostler, G. S. Denicol, J. Noronha, R. P. G. Andrade,
and F. Grassi, Phys. Rev. C 88, 044916 (2013); J. Noronha-Hostler, J. Noronha and F. Grassi,
Phys. Rev. C 90, 034907 (2014)}
\def\rosegale{J. B. Rose, J. F. Paquet, G. S. Denicol, M. Luzum, B. Schenke, S. Jeon and C. Gale, 
{\NPA{931}{926}{2014}}.}
\def\bassprl{ N.~Demir and S.~A.~Bass,
Phys. Rev. Lett. {\bf 102}, 172302 (2009).}
\def\phsdbrat{V.~Ozvenchuk, O.~Linnyk, M.~I.~Gorenstein, E.~L.~Bratkovskaya and W.~Cassing, Phys. Rev. C {\bf 87},  064903 (2013).}
\def\florkowski{W. Broniowski and W. Florkowski, {\PLB{673}{142}{2009}}.}
\def\albright{M. Albright and J.I. Kapusta,{\PRC{93}{014903}{2016}}.}
\def\hirano{P. Romatschke and U. Romatschke, Phys. Rev. Lett.{\bf 99},172301, (2007); T. Hirano and
 M. Gyulassy, Nucl. Phys. {\bf A 769}, 71, (2006).}
\def\daniel{P. Danielewicz, M. Gyulassy, {\PRD{31}{53}{1985}}.}
\def\kss{P. Kovtun, D.T. Son and A.O. Starinets, Phys. Rev. Lett.{\bf 94},
 111601, (2005).}
\def\schenke{C.Gale, S. Jeon and B. Schenke, International Journal of Modern Physics A {\bf 28}, 134011,(2013).}
\def\denicolhydro{G.S. Denicol, H. Niemi, E. Molnar and D.H. Rischke,{\PRD{85}{114047}{2012}}.}
\def\denicolpre{M. Greif, F. Reining, I. Bouras , G.S. Denicol, Z. Xu and  C. Greiner, Phys.Rev. {\bf E87} ,033019(2013).}
\def\denicolheat{G.S. Denicol, H. Niemi, I. Bouras E. Molnar , Z. Xu , D.H. Rischke, C. Greiner ,{\PRD{89}{074005}{2014}}.}
\def\rincon{J.I. Kapusta and J.M. Torres-Rincon,{\PRC{86}{054911}{2012}}.}
\def\rinconprd{J.I. Kapusta and J.M. Torres-Rincon,{\PRC{86}{054911}{2012}}.}
\def\ghoshthermal{Sabyasachi Ghosh, International Journal of Modern Physics {\bf E24},1550058,2015. }
\def\kubo{R. Kubo,J. Phys. Soc. Jpn. {\bf 12},570,(1957).}
\def\klevansky{P. Zhuang,J. Hufner, S.P. Klevansky and L. Neise,{\PRD{51}{3728}{1995}}. } 
\def\klevanskynpa{P. Rehberg, S.P. Klevansky and ,J. Hufner,{\NPA{608}{356}{1996}}. } 
\def\transqcd{P. Arnold,G.D. Moore and L.G. Yaffe,JHEP, 11, 2000, 001; ibid, JHEP 01 (2003) 030; ibid, JHEP 05 (2003) 051}
\def\quasip{M. Bluhm, B. Kampfer and K. Redlich,{\PRC{79}{055207}{2009}}.}
\def\blumredlich{M. Bluhm, B. Kamfer and K. Redlich,{\PRC{84}{025201}{2011}}.}
\def\marty{R. Marty, E. Bratkovskaya, W. Cassing, J. Aichelin and H . Berrehrah, {\PRC{88}{045204}{2013}}.}
\def\voskresenskynpaa{A.S. Khvorostukhin, V.D. Toneev and D.N. Voskresensky,{\NPA{915}{158}{2013}}.}
\def\voskresensky{A.S. Khvorostukhin, V.D. Toneev and D.N. Voskresensky,
{\NPA{845}{106}{2010}}.}
\def\gavin{Sean Gavin,{\NPA{435}{826}{1985}}.}
\def\hosoya{A. Hosoya and K. Kajantie ,{\NPB{250}{666}{1985}}.}
\def\degroot{ S.R. deGroot, W.A. van Leeuwen and Ch. G. van Weert,{\it Relativistic Kinetic Theory: Principles and Applications( North-Holland, Amsterdam, 1980)}.}
\def\lang{R. Lang, N. Kaiser, W. Weise, Eur. Phys. {\bf A48}, 109, 2012.}
\def\langweise{R. Lang, N. Kaiser, W. Weise, Eur. Phys. {\bf A50 }, 63, 2014.}
\def\ghoshkrein{Sabyasachi Ghosh, Thiago C. Peixoto, Victor Roy, Fernando E. Serna, Gastão Krein e-Print: arXiv:1507.08798 [nucl-th], (2015).}
\def\mpiexpt{K. Hagiwara {\em et al},{\PRD{66}{010001}{2002}}.}
\def\fpiexpt{B. Holostein,{\PLB{244}{83}{1990}}.}
\def\condsum{H.G. Dosch and S. Narrison,{\PLB{417}{173}{1998}}.}
\def\condlat{L. Giusti,F. Rapuano, M. Talevi and A. Vladikas,{\NPB{538}{249}{1999}}.}
\def\matiello{S. Matiello, arXiv:1210.1038[hep-ph].}
\def\nicola{ D. Fernandiz-Fraile and A. Gomez Nicola, {\EPJC{62}{37}{2009}}.}
\def\nam{ S.  Nam, Mod. Phys. Lett. A 30, 1550054,2015.}
\def\fukutome{ M. Iwasaki and T. Fukutome, J. Phys. G36, 115012, 2009.}
\def\sourav{ S. Mitra and S. Sarkar,{\PRD{89}{054013}{2014}}.}
\def\sabyath{S. Ghosh, Int.J.Mod.Phys. E24 (2015) 07, 1550058}
\def\prakash{ M. Prakash, M. Prakash, , R. Venugopalan and G. Welke,{\PR{227}{321}{1993}}.}
\def\buballarev{M. Buballa,{\PR{407}{205}{2005}}.}
\def\heckmann{K. Heckmann, M. Buballa and J. Wambach ,{\EPJA{48}{142}{2012}}}.
\def\quack{E. Quack, P. Zhuang, Y. Kalinovsky, S.P. Klevansky and
J. Hufner,{\PLB{348}{1}{1995}}}.
\def\zhuang{P. Zhuang,J. Hufner, S.P. Klevansky {\NPA{576}{525}{1994}}. } 
\def\gondolo{J. Edsjo, and P. Gondolo, {\PRD{56}{1879}{1997}}. } 
\def\goity{J. I. Goity, and H. Leutwyler, {\PLB{228}{517}{1989}}. } 
\def\page12{D. Page and S. Reddy, (2012), arXiv:1201.5602 [nucl-
th].}
\def\yakovlev{D. G. Yakovlev, A. D. Kaminker, O. Y. Gnedin, and P. Haensel, Phys. Rept. 354, 1 (2001),
 arXiv:astro- ph/0012122 [astro-ph].}
\def\kaminker{D. G. Yakovlev, O. Y. Gnedin, A. D. Kaminker, K. P.
Levenfish, and A. Y. Potekhin, Adv. Space Res. 33,
523 (2003).}
\def\rmode{N. Andersson, Astrophys. J. 502, 708 (1998), arXiv:gr-
qc/9706075 [gr-qc];
N. Andersson and K. D. Kokkotas, Mon. Not. Roy. As-
tron. Soc. 299, 1059 (1998), arXiv:gr-qc/9711088 [gr-
qc].}
\def\jharmode{ T.K. Jha, H. Mishra, V. Sreekanth,{\PRC{82}{025803}{2010}}.}
\def\kapustabk{J. I. Kapusta,{\underline Relativistic Nuclear Collisions}, Landolt-Bornstein new Series, Vol I/23,
Ed. R. Stock (Springer Verlag, Berlin Heidelberg 2010).}
\def\deb{Paramita Deb, Guru Prakash Kadam, Hiranmaya Mishra, {\PRD{94}{094002}{2016}}.}
\def\Gyulassylarry2005{M. Gyulassy and L. McLerran, {\NPA{750}{30}{2005}}.}
\def\luzumromatschke{M. Luzum and P. Romatschke, {\PRL{103}{262302}{2009}}.}
\def\dirkprl{H. Niemi, G.S. Denicol, P. Huovienen, E. Molnar and D.H. Rischke,{\PRL{106}{212302}{2011}}.}
\def\souravsuk{S. Mitra and S. Sarkar, {\PRD{87}{094026}{2013}, S. Mitra, S. Gangopadhyaya, S. Sarkar, {\PRD{91}{094012}{2015}}.}}
\def\kapustalarryprl{L. P. Csernai, J. I. Kapusta and L. D. McLerran, {\PRL{97}{152303}{2006}}.}
\def\pracheta{Pracheta Singha,Aman Abhishek, Guru Kadam, Sabyasachi Ghosh and Hiranmaya Mishra, arXiv:1705.03084v2[hep-ph].}
\def\cbmbook{B. Friman, C. Hohne, J. Knoll, S. Leupold, J. Randrup,R. Rapp. {\em et.al.}, {\em The CBM Physics Book: Compressed Baryonic
Matter in Laboratory Experiments. Lecture Notes in Physics, Springer, Berlin, Heidelberg,2011}.}
\def\jinrwhite{D. Blaschke, J. Aichelin, E. Bratkovskaya, V. Friese 
{\em et.al.}{\it Topical issues on exploring strongly interacting matter 
at high densities- nica white paper}, {\EPJA{52}{267}{2016}}.}
\def\rishi{Sreemoyee Sarkar and Rishi Sharma,  {\PRD{96}{094025}{2017}.}}
\def\shenke{C. gale, S. Jeon, B. Schenke, Int. J. Mod. Phys. {\bf A 28}, 134011,
(2013).}
\def\chamel{N. Chamel and P. Hansel, Living Rev. Rel. {\bf 11}, 10 (2008), arXiv:0812.3955[astro-ph]}
\def\sreddy{D. Page and S. Reddy,
     Annual Reviews of Nuclear and Particle Science
    {\bf  56}, 327 (2006).}
\def\bjschaefer{B. J. Schaefer, J. M. Pawlowski and J. Wambach, {\PRD{76}{074023}{2007}}.}

\def\rischkeqm{O. Scavenius, A. Mocsy, I. N. Mishustin, D. H. Rischke,{\PRC{64}{045202}{2001}}.}
\def\guptatiwari{U.S. Gupta, V.K. Tiwari,{\PRD{85}{014010}{2012}}.}
\def\bielich{B.W. Mintz, R.Stiele, R.O. Ramos, J.S. Bielich,{\PRD{87}{036004}{2013}}}
\def\ranjita{H. Mishra, R.K. Mohapatra,{\PRD{95}{094014}{2017}}.}
\def\ghoshraha{S.K. Ghosh,A. Lahiri, S. Majumder, M.G. Mustafa, S. Raha, R. Ray,{\PRD{90}{054030}{2014}}}
\def\buballa{S. Carignano, M. Buballa, W.Elkamhawy,{\PRD{94}{034023}{2016}}}
\def\pion{D.T. Son, M.A. Stephanov,{\PRD{66}{076011}{2002}};B.B.Brandt, A. Francis, H.B. Meyer, D. Robaina,{\PRD{92}{094510}{2015}}; S. Gupta, R. Sharma arXiv:1710.05345[hep-ph](2017). }
\def\dobadoestrada{A. Dobado and S.N. Santalla,{\PRD{65}{096011}{2002}}; A. Dobado, F.J. Llanes-Estrada,{\PRD{69}{116004}{2004}}}
\def\mitra{S. Mitra, S. Ghosh and S. Sarkar,{\PRC{85}{064917}{2012}}}

\end{document}